\RequirePackage{lineno}

\documentclass[preprint,onecolumn,aps,prd,groupedaddress,nofootinbib,showpacs]{revtex4}
\usepackage{graphicx}
\graphicspath{{figures/}}\DeclareGraphicsExtensions{.eps}
\usepackage{amsmath}
\usepackage{bm}
\usepackage{slashed}
\usepackage[hypertex]{hyperref}
\usepackage{xcolor}
\usepackage{amsfonts}

\allowdisplaybreaks

\newcommand{\jpsi}{{J/\psi}}

\newcommand{\state}[4]{{^#1\hspace{-0.6mm}#2_{#3}^{[#4]}}}

\newcommand\CSaSz{\state{1}{S}{0}{1}}
\newcommand\CSaPa{\state{1}{P}{1}{1}}

\newcommand\CScPz{\state{3}{P}{0}{1}}
\newcommand\CScPa{\state{3}{P}{1}{1}}
\newcommand\CScPb{\state{3}{P}{2}{1}}
\newcommand\CScPj{\state{3}{P}{J}{1}}

\newcommand\COaPa{\state{1}{P}{1}{8}}

\newcommand\COcPz{\state{3}{P}{0}{8}}
\newcommand\COcPa{\state{3}{P}{1}{8}}
\newcommand\COcPb{\state{3}{P}{2}{8}}

\newcommand{\initialstate}[2]{{#1^{[#2]}}}
\newcommand\vone{\initialstate{v}{1}}
\newcommand\veight{\initialstate{v}{8}}
\newcommand\aone{\initialstate{a}{1}}
\newcommand\aeight{\initialstate{a}{8}}
\newcommand\tone{\initialstate{t}{1}}
\newcommand\teight{\initialstate{t}{8}}

\newcommand\DeltaZero{\Delta_0}
\newcommand\DeltaPPZero{{\Delta_0''}}
\newcommand\DPlusOne{\Delta_+^{[1]}}
\newcommand\DPlusPOne{{{\Delta_+^{[1]}}'}}
\newcommand\DPlusPPOne{{{\Delta_+^{[1]}}''}}
\newcommand\DPlusEight{\Delta_-^{[8]}}
\newcommand\DPlusPEight{{{\Delta_-^{[8]}}'}}
\newcommand\DPlusPPEight{{{\Delta_-^{[8]}}''}}
\newcommand\DMinusOne{\Delta_-^{[1]}}
\newcommand\DMinusPOne{{{\Delta_-^{[1]}}'}}
\newcommand\DMinusPPOne{{{\Delta_-^{[1]}}''}}
\newcommand\DMinusEight{\Delta_+^{[8]}}
\newcommand\DMinusPEight{{{\Delta_+^{[8]}}'}}
\newcommand\DMinusPPEight{{{\Delta_+^{[8]}}''}}
\newcommand\DPMOne{\Delta_\pm^{[1]}}
\newcommand\DPMEight{\Delta_\pm^{[8]}}
\newcommand\DPMPOne{{{\Delta_\pm^{[1]}}'}}
\newcommand\DPMPEight{{{\Delta_\pm^{[8]}}'}}
\newcommand\DPMPPOne{{{\Delta_\pm^{[1]}}''}}
\newcommand\DPMPPEight{{{\Delta_\pm^{[8]}}''}}
\newcommand\DAA{{\delta(1-z+\zeta_1)}}
\newcommand\DBB{{\delta(1-z-\zeta_1)}}
\newcommand\DXX{{\delta(1-z+\zeta_2)}}
\newcommand\DYY{{\delta(1-z-\zeta_2)}}
\newcommand\DPAA{{\delta'(1-z+\zeta_1)}}
\newcommand\DPBB{{\delta'(1-z-\zeta_1)}}
\newcommand\DPXX{{\delta'(1-z+\zeta_2)}}
\newcommand\DPYY{{\delta'(1-z-\zeta_2)}}

\newcommand\as{\alpha_s}

\newcommand\LogUV{\,\text{ln}\big[\frac{\mu_0^2}{m_Q^2}\big]}
\newcommand\LogIR{\,\text{ln}\big[\frac{\mu_\Lambda^2}{m_Q^2}\big]}
\newcommand{\mylog}[1]{{\,\text{ln}(#1)}}
\newcommand\logtwo{\,\text{ln}\, 2}

\newcommand{\Plusz}[1]{{\left( #1 \right)_{0+}}}
\newcommand{\Minusz}[1]{{\left( #1 \right)_{0-}}}
\newcommand\Plusa{\left(\frac{1}{\zeta_1}\right)_{1+}}
\newcommand\Plusb{\left(\frac{1}{\zeta_1^2}\right)_{2+}}

\newcommand\Minusa{\left(\frac{1}{\zeta_1}\right)_{1-}}
\newcommand\Minusb{\left(\frac{1}{\zeta_1^2}\right)_{2-}}
\newcommand\Minusc{\left(\frac{1}{\zeta_1^3}\right)_{3-}}
\newcommand\PlusLoga{\left(\frac{\text{ln}(\zeta_1^2)}{\zeta_1}\right)_{1+}}

\newcommand\MinusLogb{\left(\frac{\text{ln}(\zeta_1^2)}{\zeta_1^2}\right)_{2-}}

\newcommand\cc{Q\bar{Q}}

\newcommand{\ben}{\begin{eqnarray}}
\newcommand{\een}{\end{eqnarray}}

\newcommand{\bef}{\begin{figure}[!htp]}
\newcommand{\eef}{\end{figure}}

\newcommand{\bea}{\begin{eqnarray}}
\newcommand{\eea}{\end{eqnarray}}

\newcommand{\Swave}{Swave}

\begin{document}

\title{Heavy quarkonium fragmentation functions from a heavy quark pair. II. $P$ wave}

\author{Yan-Qing Ma}
\email{yqma@bnl.gov}
\affiliation{Physics Department, Brookhaven
National Laboratory, Upton, NY 11973, USA}

\author{Jian-Wei Qiu}
\email{jqiu@bnl.gov}
\affiliation{Physics Department, Brookhaven
National Laboratory, Upton, NY 11973, USA}
\affiliation{C.N. Yang
Institute for Theoretical Physics and Department of Physics and Astronomy, Stony Brook University, Stony
Brook, NY 11794, USA}

\author{Hong Zhang}
\email{hong.zhang@stonybrook.edu}
\affiliation{Department of Physics and Astronomy, Stony Brook University, Stony Brook, NY 11794, USA}

\date{\today}
\preprint{YITP-SB-13-50}
\begin{abstract}

Recently, a new perturbative QCD factorization formalism for heavy quarkonium production
at a large transverse momentum was proposed. Phenomenological application of this new approach
relies on our knowledge of a large number of universal fragmentation functions (FFs)
at an input factorization scale $\mu_0\gtrsim 2m_Q$ with heavy quark mass $m_Q$,
which are nonperturbative, and in principle, should be extracted from data.
With heavy quark mass $m_Q\gg \Lambda_{\rm QCD}$, we calculate these input FFs
in terms of non-relativistic QCD (NRQCD) factorization.  
We derived contributions to these input FFs through all $S$-wave NRQCD $Q\bar{Q}$-states
in a companion paper \cite{\Swave}.  
In this paper, we calculate contributions to the heavy quark-pair FFs from
all $P$-wave NRQCD $Q\bar{Q}$-states.

\end{abstract}


\maketitle

%%%%%%%%%%%%%%%%%%%%%%%%%%%%%%%%%
%                                                   						  %
%                                   Introduction						  %
%  											   	  %
%%%%%%%%%%%%%%%%%%%%%%%%%%%%%%%%%
\section{introduction}

Studying the mechanism of heavy quarkonium production is important for understanding QCD and the strong interaction dynamics \cite{Brambilla:2010cs,Bodwin:2013nua}.  Over the years, theory calculations of heavy quarkonium production relied on various models of the production mechanism, and have had successes and failures in interpreting existing data \cite{Abulencia:2007us,Chatrchyan:2012woa}.
Heavy quarkonium production is still puzzling us after almost forty years since the discovery of $J/\psi$ \cite{Aubert:1974js, Augustin:1974xw}.
Recently, a systematic perturbative QCD (pQCD) factorization approach to high $p_T$ heavy quarkonium production at collider energies was proposed \cite{KMQS-hq1,KMQS-hq2}, following some earlier works \cite{Nayak:2005rw,Nayak:2005rt,Kang:2011mg,Kang:2011zza}.  A similar factorization approach based on soft-collinear effective theory (SCET) was also recently proposed \cite{Fleming:2012wy}. With the advances in theory and tremendous amount of precise data from the LHC, it is an excellent time to study the physics of heavy quarkonia to resolve the long-standing puzzles associated with heavy quarkonium production.

The pQCD factorization approach first expands the cross section of heavy quarkonium production at large transverse momentum $p_T$ in terms of powers of $1/p_T$, and then, factorizes the leading power (LP) and next-to-leading power (NLP) contributions to the cross section in terms of perturbatively calculable short-distance partionic hard parts and long-distance nonperturbative, but universal fragmentation functions (FFs) and parton distribution functions (PDFs) \cite{KMQS-hq1}. Knowing the universal PDFs and perturbatively calculated partonic hard parts \cite{KMQS-hq2}, the predictive power of the pQCD factorization formalism relies on our knowledge of many FFs, more precisely, a minimum of eight to ten FFs, depending on how many single heavy flavor fragmentation channels are included, for each heavy quarkonium state produced \cite{\Swave}.
With perturbatively calculated evolution kernels of FFs \cite{KMQS-hq1}, it is the FFs at an initial factorization scale $\mu_0\gtrsim 2m_Q$ with heavy quark mass $m_Q$ that are needed.  These input FFs are non-perturbative, and in principle, should be extracted from experimental data.  However, in practice,
it is hard to extract so many input FFs from data directly, and thus it is difficult to test this pQCD factorization formalism precisely.

Perturbatively calculated partonic hard parts and evolution kernels of FFs are the same for all heavy quarkonia produced.  It is the input FFs that are sensitive to the individual properties of each heavy quarkonium state produced, including its spin and polarization.  Although these input FFs are non-perturbative in terms of pQCD
factorization approach, they are different from the better-known FFs to light hadrons because of the large heavy quark mass $m_Q\gg \Lambda_{\rm QCD}$.  With the large heavy quark mass and the fact that $\mu_0 \gtrsim 2m_Q$, it might be possible to calculate these input FFs by using the non-relativistic QCD (NRQCD)
\cite{Bodwin:1994jh}, an effective theory of QCD.   
We are aware that without a formal proof of NRQCD factorization for calculating these input FFs, some modifications to these calculated input FFs might be needed for a better description of data.  However,
knowing the phenomenological successes of NRQCD factorization approach to heavy quarkonium production \cite{Brambilla:2010cs, Kramer:2001hh, Braaten:1996pv, Petrelli:1997ge}, we are confident that applying NRQCD to the input FFs, we should, at least, be able to derive reasonable constraints on the functional form of these input FFs, in terms of a very limited number of NRQCD long-distance matrix elements (LDMEs) with all coefficients - the FFs' functional dependence on parton momentum fractions derived by matching between pQCD and NRQCD.

In our companion paper \cite{\Swave}, we calculated contributions to the input FFs for a perturbatively produced heavy quark pair to fragment into a physical quarkonium via all intermediate $S$-wave NRQCD $Q\bar{Q}$-states, which are the most important contribution to the FFs from NRQCD factorization. However, contributions to heavy quarkonium production from $P$-wave NRQCD $Q\bar{Q}$-states are also very important, and sometimes, indispensable. First of all, $P$-wave quarkonium production, such as $\chi_{cJ}$, must have a major contribution through $P$-wave NRQCD $Q\bar{Q}$-states.  Second, $S$-wave quarkonium production, such as $\jpsi$, usually receives significant feed-down contribution from decay of $P$-wave quarkonia. In addition, from the NRQCD factorization approach, direct production of $S$-wave quarkonium can also receive relativistic correction from $P$-wave color-octet NRQCD $Q\bar{Q}$-states.  In this paper, we calculate, up to next-to-leading order (NLO) in $\alpha_s$ expansion, the contributions to the same input FFs from all intermediate $P$-wave NRQCD $Q\bar{Q}$-states.

We should point out that NLO calculation in NRQCD factorization approach through $P$-wave NRQCD $Q\bar{Q}$-states are available for heavy quarkonium production in hadron-hadron collisions \cite{Ma:2010vd,Ma:2010yw,Butenschoen:2010rq,Wang:2012is,Gong:2013qka}, electron-hadron collisions \cite{Butenschoen:2009zy}, and electron-position collisions \cite{Zhang:2009ym}.  From our results in this paper, the evolution kernels of FFs in Ref.~\cite{KMQS-hq1} and partonic hard parts available in Ref.~\cite{KMQS-hq2}, we can in principle reproduce large $p_T$ or high energy results of the previous NLO NRQCD calculations and more, including the resummation of large logarithms from the evolution of the FFs.  The numerical effort to solve for the FFs from our calculated input FFs and the calculated evolution kernels, and to combine the evolved FFs and available partonic hard parts to predict the cross sections for heavy quarkonium production is beyond the scope of this paper, and we leave it for the future publication.

The rest of this paper is organized as follows. In section \ref{sec:NRQCD}, we briefly review the definition of the $Q\bar{Q}$-pair FFs, and our approach to calculate these FFs at an input factorization scale $\mu_0$ in terms of NRQCD factorization formalism.  In section \ref{sec:general}, we introduce necessary symmetries that help simplify our calculations, and discuss how can the charge conjugate symmetry and parity conservation constrain the structure of these input FFs.  In addition, we give a detailed discussion of the Coulomb singularity in our calculation in appendix \ref{app:coulomb} to justify our method: expanding the relative momentum between the final-state heavy quark pair before doing loop integral of the relative momentum of initial-state heavy quark pair.  We present LO and NLO calculation of the FFs with some explicit examples in Section \ref{sec:LO} and \ref{sec:NLO}, respectively.  Our complete results for input FFs of a perturbatively produced heavy quark pair to fragment into a physical quarkonium through all relevant $P$-wave NRQCD $Q\bar{Q}$-states are listed in appendix \ref{app:results}.  Our conclusions are summarized in Section \ref{sec:summary}.

%%%%%%%%%%%%%%%%%%%%%%%%%%%%%%%%%
%                                                   						  %
%                                   framework						  %
%  											   	  %
%%%%%%%%%%%%%%%%%%%%%%%%%%%%%%%%%
\section{Fragmentation functions for a perturbative heavy quark pair to fragment into a $P$-wave NRQCD state}
\label{sec:NRQCD}

The FF for a perturbatively created $Q\bar{Q}$-pair of spin-color quantum number $\kappa$ to fragment into a physical heavy quarkonium $H$ is defined as~\cite{KMQS-hq1}
%==========================
\begin{linenomath*}
\begin{align}\label{eq:QQFF}
\begin{split}
{\cal D}_{[\cc(\kappa)]\to H}&(z,\zeta_1,\zeta_2;m_Q,\mu_0)
\hspace{-0.1cm}
=
\hspace{-0.2cm}
\int\frac{p^+ dy^-}{2\pi}\frac{p^+ \hspace{-0.1cm} /z dy_1^-}{2\pi}\frac{p^+ \hspace{-0.1cm} /z dy_2^-}{2\pi}
e^{-i(p^+ \hspace{-0.1cm} /z)y^-}e^{i(p^+ \hspace{-0.1cm} /z)[(1-\zeta_2)/2]y_1^-}e^{-i(p^+ \hspace{-0.1cm} /z)[(1-\zeta_1)/2]y_2^-}\\
&
\times {\cal P}_{ij,kl}^{(s)}(p_c){\cal C}_{ab,cd}^{[I]}
\langle 0 |\bar{\psi}_{c',k}(y_1^-) [\Phi_{\hat{n}}^{(F)}(y_1^-)]_{c'c}^\dag [\Phi_{\hat{n}}^{(F)}(0)]_{d d'}\, \psi_{d',l}(0) |H(p) X\rangle\\
&\times
\langle H(p)X|
\bar{\psi}_{a',i}(y^-) [\Phi_{\hat{n}}^{(F)}(y^-)]^\dag_{a' a} [\Phi_{\hat{n}}^{(F)}(y^-+y_2^-)]_{b b'} \psi_{b',j}(y^-+y_2^-) |0\rangle,
\end{split}
\end{align}
\end{linenomath*}
%==========================
where
operators ${\cal P}_{ij,kl}^{(s)}(p)$ and ${\cal C}_{ab,cd}^{[I]}$ project the initial $\cc$-pair to a definite spin and color state $\kappa$, which could be a vector ($v^{[1,8]}$), axial-vector ($a^{[1,8]}$) or tensor ($t^{[1,8]}$) state, with superscript $^{[1]}$ ($^{[8]}$) denoting a color singlet (color octet) state.
Definitions of these projection operators
could be found in Refs.~\cite{KMQS-hq1,\Swave}.
Subscripts $i,j,k,l$ ($a,a',b,b'\ldots$) are spin (color) indices, with summation over repeated indices understood.
$z$ is the light-cone momentum fraction of the quarkonium $H$ with respect to the momentum of initial fragmenting $Q\bar{Q}$-pair, $p_c$.
$\zeta_1$ ($\zeta_2$) is the relative momentum fraction of the fragmenting $\cc$-pair in the amplitude (the complex conjugate of the amplitude).
Although the total momentum $p_c$ of the $\cc$-pair in the amplitude and its complex conjugate must be the same, the relative momenta, denoted by $\zeta_1$ and $\zeta_2$, could be different.
$\Phi_{\hat{n}}^{(F)}$ in Eq.~(\ref{eq:QQFF}) is a gauge link along the $\hat{n}$ direction,
which keeps the definition in Eq.~\eqref{eq:QQFF} gauge invariant, defined as
%==========================
\begin{linenomath*}
\begin{align}\label{eq:GLdef}
\begin{split}
\Phi_{\hat{n}}^{(F)}(y^-)={\cal P}\,\text{exp}\left[-i\,g \int_{y^-}^{\infty} d\lambda \,{\hat{n}}\cdot A^{(F)}(\lambda {\hat{n}})\right],
\end{split}
\end{align}
\end{linenomath*}
%==========================
where $\mathcal{P}$ denotes path ordering and the superscript $(F)$ represents fundamental color representation.  The superscripts ``$+$'' and ``$-$'' in above equations represent the light-cone components of any four-vector, $x^\mu=(x^0,x^1,x^2,x^3)$, as $x^{\pm} = (x^0\pm x^3)/\sqrt{2}$. Specifically, $\hat{n}^\mu = (0^+, 1^-, 0_\perp)$ and $p\cdot\hat{n}=p^+$.

Applying NRQCD factorization~\cite{Bodwin:1994jh} to the $\cc$-pair FF in Eq.~\eqref{eq:QQFF}, we could express the FF as a sum of infinite terms, each of which is factorized into a product of a short-distance coefficient and a NRQCD LDME \cite{Kang:2011zza,KMQS-hq2},
%=========================
\begin{linenomath*}
\begin{align}\label{eq:QQFFNR1}
\begin{split}
{\cal D}_{[\cc(\kappa)]\to H}(z,\zeta_1,\zeta_2; m_Q,\mu_0)
=
\sum_{[\cc(n)]} \hat{d}_{[\cc(\kappa)]\to[\cc(n)]}(z,\zeta_1,\zeta_2;m_Q,\mu_0,\mu_\Lambda)
\langle \mathcal{O}_{[\cc(n)]}^{H}(\mu_\Lambda)\rangle,
\end{split}
\end{align}
\end{linenomath*}
%=========================
where $\mu_0$ and $\mu_\Lambda$ are pQCD and NRQCD factorization scales, respectively.
The summation of intermediate $[\cc(n)]$ runs over all possible NRQCD states, which are labelled by spectroscopic notation $\state{{2S+1}}{L}{J}{1,8}$.
Short-distance coefficients describe physics with energy scale larger than $\mu_\Lambda \gg \Lambda_{\text{QCD}}$, which can be calculated perturbatively, while LDMEs represent the physics with energy scale smaller than $\mu_\Lambda$, which need to be determined by fitting experimental data. To calculate these short-distance coefficients $\hat{d}_{[\cc(\kappa)]\to[\cc(n)]}$ order by order in $\as$, we replace the quarkonium state $H$ in Eq.~(\ref{eq:QQFFNR1}) by some asymptotic NRQCD states, $[\cc(n')]$, since the short-distance coefficients are insensitive to the details of the heavy quarkonium produced,
%=========================
\begin{linenomath*}
\begin{align}\label{eq:QQFFNR2}
\begin{split}
{\cal D}_{[\cc(\kappa)]\to [\cc(n')]}(z,\zeta_1,\zeta_2; m_Q,\mu_0)
=
\hspace{-0.3cm}
\sum_{[\cc(n)]}  \hat{d}_{[\cc(\kappa)]\to[\cc(n)]}(z,\zeta_1,\zeta_2;m_Q,\mu_0,\mu_\Lambda)
\langle \mathcal{O}_{[\cc(n)]}^{[\cc(n')]}(\mu_\Lambda)\rangle.
\end{split}
\end{align}
\end{linenomath*}
%=========================
The left-hand-side (LHS) of Eq.~\eqref{eq:QQFFNR2} can be calculated perturbatively in perturbative QCD with a NRQCD projection for the state $[\cc(n')]$, while $\langle \mathcal{O}_{[\cc(n)]}^{[\cc(n')]}(\mu_\Lambda)\rangle$ on the right-hand-side (RHS) can be calculated perturbatively in the NRQCD.  With the calculated LHS and RHS, we can extract all short-distance coefficients, $\hat{d}_{[\cc(\kappa)]\to[\cc(n)]}$, order-by-order in power of $\as$, from the matching condition of Eq.~\eqref{eq:QQFFNR2}, if the conjecture of NRQCD factorization is actually valid.
Both the LHS and RHS could be infrared (IR) divergent and Coulomb divergent.  But, all these divergences should be cancelled order by order between LHS and RHS, and leave the short-distance hard parts IR safe if the NRQCD factorization in Eq. (3) is valid.
As we show in this paper by explicit calculations, up to NLO, all $\hat{d}_{[\cc(\kappa)]\to[\cc(n)]}$ are indeed finite. However, the lack of an all-order proof of NRQCD factorization still leaves some doubts on if such beautiful cancellation of IR and Coulomb divergences could be true at higher orders.

Since short-distance coefficients are derived following the cancelation of IR and Coulomb divergences, it is necessary to introduce some kind of regulators to regularize the singularities.  In this paper, like what we did in our companion paper \cite{\Swave}, we adopt the dimensional regularization.
The LHS of Eq.~(\ref{eq:QQFFNR2}) in the $D$-dimension can be written as
%==========================
\begin{linenomath*}
\begin{align}\label{eq:QQFFQQ}
\begin{split}
{\cal D}_{[\cc(s^{[b]})]\to [\cc(i^{[b']})]}
=&
\frac{z^{D-2}}{N_s N_b N_{i}^\text{NR} N_{b'}^\text{NR}}
\int\frac{{\text d}^{D} p_c}{(2\pi)^{D}}
\left(\prod_X\int\frac{{\text d}^{D-1} p_X}{(2\pi)^{D-1} 2E_X}\right)
\delta\left(z-\frac{p^+}{p_c^+}\right)
\,\\
&
\times
(2\pi)^D\delta^D(p_c-p-\sum_X p_X)
{\cal M}_{[\cc(s^{[b]})]\to [\cc(i^{[b']})]}(p, z, \zeta_1, \zeta_2)\\
=&
\frac{z^{D-2}}{N_s N_b  N_{i}^\text{NR} N_{b'}^\text{NR}}
\left(\prod_X\int\frac{{\text d}^{D-1} p_X}{(2\pi)^{D-1} 2E_X}\right)
\delta\left(z-\frac{p^+}{p_c^+}\right)
\\
&\times
{\cal M}_{[\cc(s^{[b]})]\to [\cc(i^{[b']})]}(p, z, \zeta_1, \zeta_2)\, ,
\end{split}
\end{align}
\end{linenomath*}
%==========================
where we have separated the spin and color labels for the initial- and final-state $\cc$-pair,
with $s$ and $i$ for spin and $b$ and $b'$ for color, respectively.
The $N$'s are different normalization factors for spin and color,
given in Appendix A of our companion paper \cite{Swave}.
The matrix element $\mathcal{M}$ has the explicit form
%==========================
\begin{linenomath*}
\begin{align}\label{eq:QQFFQQM2}
\begin{split}
{\cal M}_{[\cc(s^{[b]})]\to [\cc(i^{[b']})]}(p, z, \zeta_1, \zeta_2)
=&\,
{\text{Tr}}\left[\Gamma_s(p_c)\, C_b\, {\cal A}_{[\cc(s^{[b]})]\to [\cc(i^{[b']})]}(p, z, \zeta_1)\right] \\
&
\hspace{-1.5cm}\times
{\text{Tr}}\left[\Gamma_s^\dag(p_c)\, C_b^\dag\, {\cal A}^\dag_{[\cc(s^{[b]})]\to \cc[i^{[b']}]}(p, z,\zeta_2)\right]
\times P_s(p_c)\, P_{i}^\text{NR}(p)
\, ,
\end{split}
\end{align}
\end{linenomath*}
%==========================
where ``Tr" denotes the trace for both color and $\gamma$-matrices.
$\Gamma_s$ and $C_b$ are spin and color projection operators for initial $\cc$-pair.
$P_s$ ($P_{i}^{\text{NR}}$) is the summation of polarizations, i.e. $\Sigma_{\lambda}\epsilon^*_\lambda(p_c)  \epsilon_\lambda(p_c)$ ($\Sigma_{\lambda'}\epsilon^*_{\lambda'}(p)  \epsilon_{\lambda'}(p)$) for initial (final) $\cc$-pair.
The amplitude $\mathcal{A}$ could be calculated by
%==========================
\begin{linenomath*}
\begin{align}\label{eq:QQFFQQA}
\begin{split}
{\cal A}_{[\cc(s^{[b]})]\to [\cc(i^{[b']})]}(p,z,\zeta_1)
=&
\lim_{q_r\to0}\left(\prod_{j=0}^L
\frac{\text{d}}{\text{d}{q_r^{\alpha_j}}}\right)
\bigg\{
\int\frac{{\text d}^D q_1}{(2\pi)^D}
\ 2\ \delta(\zeta_1-\frac{2q_1^+}{p_c^+})\,
\\
&
\qquad\qquad\times
{\cal \bar{A}}_{[\cc(s^{[b]})]\to [\cc(i^{[b']})]}(q_1,q_r)\,
\Gamma_{i}^{\text{NR}}(p)\, C_{b'}^\text{NR}
\bigg\}
\, ,
\end{split}
\end{align}
\end{linenomath*}
%==========================
where $\mathcal{\bar{A}}$ is the amputated amplitude.
$q_1$ ($q_r$) is half of the relative momentum between the heavy $Q$-quark and
$\bar{Q}$-quark for the initial (final) $\cc$-pair.
$\Gamma_{i}^{\text{NR}}$ and $C_{b'}^{\text{NR}}$ are spin and color projection operators,
respectively, for final non-relativistic (NR) $\cc$-pair.
The definitions of these projection operators and normalization factors are all given in Appendix A
of our companion paper \cite{\Swave}.

We emphasize that in Eq. (\ref{eq:QQFFQQA}), the limit $q_r\to 0$ and
the derivative operations are outside of the $q_1$-integral.
However, the integration of $q_1$ with $q_r\neq 0$ is difficult and tedious.
A widely-used trick for previous NRQCD calculations of heavy quarkonium production cross sections
is to switch the $q_1$-integration with the derivative operations and the limit of $q_r\to 0$.
The validity of this trick for the cross section calculations was justified up to NLO~\cite{Beneke:1997zp}.
However, the existing proof does not directly apply to our case of $\cc$-pair FFs,
because of the $\delta$-function in Eq.~(\ref{eq:QQFFQQA}).
For producing final-state $\cc$-pair in $P$-wave, the $q_r$-derivative further complicates the situation.
After considerable algebra, we proved explicitly that
such trick to switch the $q_1$-integration and the limit of $q_r\to 0$ and the derivatives in Eq. (\ref{eq:QQFFQQA}) is still valid to NLO for producing both $S$-wave and $P$-wave final-state $\cc$-pairs.
We present our justification in Appendix \ref{app:coulomb}.

In the next three sections, we present explicit LO and NLO calculation of short-distance coefficients in NRQCD factorization approaches to the FFs, making use of Eqs.~(\ref{eq:QQFFQQ})-(\ref{eq:QQFFQQA}).  We start with some general discussions on how to use the symmetries to simplify our calculations, as well as to derive some constraints/relations between various pieces of contributions.  We emphasize that the symmetries are important for understanding the general structure of our results.

%%%%%%%%%%%%%%%%%%%%%%%%%%%%%%%%%
%                                                   						  %
%                                   LOcalculation					  %
%  											   	  %
%%%%%%%%%%%%%%%%%%%%%%%%%%%%%%%%%
\section{Symmetries} \label{sec:general}

In this section, we show how fundamental symmetries constrain the structure of the FFs calculated in NRQCD factorization approach.

\newcommand{\states}[3]{{^#1\hspace{-0.6mm}#2_{#3}}}

\subsection{Color charge conservation} \label{sec:color}

Color charge conservation could be a serious constraint for partonic contributions to $Q\bar{Q}$-pair FFs
to a non-relativistic $\cc$-pair without radiating any additional partons into the final-state.
For these FFs, such as LO contribution,
${\cal D}^{\text{LO}}_{[\cc(s^{[b]})]\to [\cc(i^{[b']})]}$, and the NLO virtual contribution,
${\cal D}^{\text{NLO-V}}_{[\cc(s^{[b]})]\to [\cc(i^{[b']})]}$, the color of the fragmenting pQCD
heavy quark pair $[\cc(s^{[b]})]$ should be the same as that of final-state non-relativistic
heavy quark pair $[\cc(i^{[b']})]$, or $b=b'$.  Due to the color normalizaiton for NRQCD matrix elements,
as defined in Appendix A of Ref.~\cite{\Swave},
color charge conservation requires ${\cal D}^{\text{LO}}_{[\cc(s^{[8]})]\to [\cc(i^{[8]})]}
=(N_c^2-1)^{-1}\times{\cal D}^{\text{LO}}_{[\cc(s^{[1]})]\to [\cc(i^{[1]})]}$.

\subsection{Lorentz invariance}\label{sec:lorentz}

Even if the initial and the final $\cc$-pair are in the same color state, partonic contributions to ${\cal D}_{[\cc(s^{[b]})]\to [\cc(i^{[b]})]}$ without radiating any parton to the final-state may still vanish, due to the Lorentz invariance, or more precisely, the angular momentum conservation.
For initial-state $s=v, a, t$ and final-state $i=\states{1}{S}{0}, \states{3}{S}{1}, \states{1}{P}{1}, \states{3}{P}{0}, \states{3}{P}{1}, \states{3}{P}{2}$, we could have a total of 18 (or 24) channels (if we distinguish the two initial tensor states).  By applying Lorentz invariance,  8 out of the 18 partonic fragmentation channels vanish.  Once all loop integrations are performed, contributions to all these fragmentation channels could only depend on two momentum vectors, $\hat{n}$ and $p$, and three polarization vectors: $\epsilon_\alpha$ for $L=1$ states,  $\epsilon_\beta$ for $S=1$ states, and $\epsilon_\rho$ if the initial $\cc$-pair is in the $t$ state.

If there is one $\gamma^5$ in the combined initial- and final-state spin projector: $\Gamma_s \Gamma_i^{\text{NR}}$, we need two of three possible polarization vectors ($\epsilon$'s discussed above) plus the two linear momenta $\hat{n}$ and $p$ to construct the Levi-Civita tensor. Consequently, the partonic fragmentation channels: $v \to \states{1}{S}{0}$, $t \to \states{1}{S}{0}$, $a \to \states{3}{S}{1}$, and $v \to \states{1}{P}{1}$ must vanish.

Since $p^\alpha$ and $p^\beta$ give zero when contracting with $\epsilon_\alpha$ (for $L=1$ states) and $\epsilon_\beta$ (for $S=1$ states), respectively, Lorentz structure of the amplitude of the process $v \to \states{3}{P}{J}$ must be a linear combination of $\hat{n}^\alpha \hat{n}^\beta$ and $g^{\alpha \beta}$, which is symmetric under the exchange of $\alpha$ and $\beta$. For the amplitude of the process $a \to \states{3}{P}{J}$, the Lorentz structure must be $\epsilon^{\alpha \beta\mu\nu}{n}_\mu p_\nu$, which is anti-symmetric under the exchange of $\alpha$ and $\beta$.
Therefore, the partonic fragmentation processes: $v \to \states{3}{P}{1}$, $a \to \states{3}{P}{0}$ and $a \to \states{3}{P}{2}$ are not allowed, since the $\states{3}{P}{0, 2}$ are symmetric between spin and orbital angular momentum while $\states{3}{P}{1}$ is antisymmetric between spin and orbital angular momentum.

Finally, the partonic fragmentation channel: $t \to \states{3}{P}{0}$ must vanish because $p^\rho$ and $\hat{n}^\rho$ give zero when contracting with the tensor polarization vector $\epsilon_\rho$, and there is no other Lorentz structure to take the index $\rho$.  Our explicit calculations up to NLO in $\as$ support our analysis and confirm these constraints.

\subsection{Reality and symmetries}\label{sec:charge}

As both the cross section and the partonic hard part are real, the heavy quark pair FFs defined in Eq.~(\ref{eq:QQFF}) is also real, ${\cal D}_{[\cc(\kappa)]\to H}(z,\zeta_1,\zeta_2;m_Q,\mu_0)^* =  {\cal D}_{[\cc(\kappa)]\to H}(z,\zeta_1,\zeta_2;m_Q,\mu_0)$. The reality requires
that these FFs are symmetric in $\zeta_1$ and $\zeta_2$.

QCD is invariant under the charge conjugation, parity, and time-reversal transformation.  But, it is not easy to apply these symmetry transformations to the FFs directly.  However, they could be used to study the symmetry properties of the matrix elements defining the FFs in Eq.~(\ref{eq:QQFF}).
Since time-reversal transformation is not unitary, its operation connects matrix elements of the states with and without time-reversal transformation \cite{Qiu:1998ia,Kang:2008ey}
\begin{eqnarray}
\langle 0 |\, \widehat{\cal O}(\psi, A_\mu)\,|H(p)X\rangle
=\left(\langle H(p)X |{\cal T}^{-1}\right)
{\cal T}   \widehat{\cal O}(\psi, A_\mu)^\dagger {\cal T}^{-1}
\left( {\cal T}|0\rangle \right)\,
\label{eq:time-r}
\end{eqnarray}
where $\widehat{\cal O}(\psi, A_\mu)$ is an operator of quark and gluon field,
${\cal T}$ is the time-reversal operator and $\left(\langle H(p)X |{\cal T}^{-1}\right)$
and $\left( {\cal T}|0\rangle \right)$ are time-reversal transformed states.
Since charge conjugation ${\cal C}$ and parity ${\cal P}$ transformation are unitary,
they can be directly inserted into the matrix element as
\begin{eqnarray}
\langle 0 |\, \widehat{\cal O}(\psi, A_\mu)\,|H(p)X\rangle
&=&
\langle 0 |\, \widehat{\cal O}(\psi, A_\mu)\,{\cal C}^{-1}{\cal C}\, |H(p)X\rangle
\nonumber \\
&=&
\langle 0 |\, \widehat{\cal O}(\psi, A_\mu)\,{\cal P}^{-1}{\cal P}\, |H(p)X\rangle\, .
\label{eq:cp}
\end{eqnarray}
For an example, applying the parity and time-reversal invariance to the matrix elements of
the FFs to a unpolarized final-state heavy quarkonium, defined in Eq.~(\ref{eq:QQFF}),
one can derive the same $\zeta_1\leftrightarrow \zeta_2$ symmetry property of the FFs
obtained by applying the reality of the FFs.

Although the charge conjugation operation ${\mathcal{C}}$ can not be applied to the FFs directly,
because the initial fragmenting $\cc$-pair is not an eigenstate of ${\cal C}$ due to its relative momentum,
we find that the FFs are actually invariant under a modified charge conjugation $\overline{\mathcal{C}}$,
if both the initial and the final heavy quark pairs are color singlet.
The modified charge conjugation operation $\overline{\mathcal{C}}$ is defined as
the charge conjugation operation ${\mathcal{C}}$ followed by reversing the direction of
the relative light cone momentum of the pair, i.e. $\zeta_1\to-\zeta_1$ for the amplitude.
More specifically, for the fragmentation from a pQCD $\cc$-pair to a non-relativistic $\cc$-pair,
the $\overline{\mathcal{C}}$ operation leads to $(-1)^{\delta_{s,a}+1}$ for the initial $\cc$-pair
with $s=v,a,t$, and $(-1)^{L+S}$ for a final non-relativistic $\cc$-pair $(^{2S+1}L_J)$.
By applying $\overline{\mathcal{C}}$ on the amplitude and keeping the complex conjugate of the amplitude untouched for a FF, one picks up an overall factor $(-1)^{L+S+\delta_{s,a}+1}$.
If there is a gluon radiated into the final-state, we can still apply the $\overline{\mathcal{C}}$ operation
as long as one of the initial and final $\cc$-pairs is in a color singlet state.
By applying $\overline{\mathcal{C}}$ operation on the amplitude of the FFs
and keeping the complex conjugate of the amplitude untouched, one picks up an overall factor
$(-1)^{L+S+\delta_{s,a}}$.

More generally, if we apply $\overline{\mathcal{C}}$ operation to both the amplitude
and its complex conjugate for heavy quark pair FFs and combine the reality, we have
%============================
\begin{align}
{\cal D}_{[\cc(s^{[b]})]\to [\cc(\state{{2S+1}}{L}{J}{b'})]}(z,-\zeta_1,-\zeta_2)
={\cal D}_{[\cc(s^{[b]})]\to [\cc(\state{{2S+1}}{L}{J}{b'})]}(z,\zeta_1,\zeta_2),
\label{eq:parity23}
\end{align}
where $b,b'=[1],[8]$.  Combining the symmetry property of the FFs when
$\zeta_1\leftrightarrow \zeta_2$ and that in Eq.~(\ref{eq:parity23}), the FFs also have
the following crossing symmetry,
%============================
\begin{align}
{\cal D}_{[\cc(s^{[b]})]\to [\cc(\state{{2S+1}}{L}{J}{b'})]}(z,-\zeta_1,-\zeta_2)
={\cal D}_{[\cc(s^{[b]})]\to [\cc(\state{{2S+1}}{L}{J}{b'})]}(z,\zeta_2,\zeta_1).
\label{eq:parity22}
\end{align}
All these symmetry properties of the FFs are verified by our explicit calculations below.

Charge conjugation could also be employed
to constrain the delta function structure of real gluon emission subprocess in our NLO calculation, for which the Feynman diagrams
are shown in Fig.~\ref{fig:NLOreal}. The analysis is easier in light cone gauge $A^+=0$, although the conclusion is gauge independent. In the light cone gauge, only the first two diagrams in Fig.~\ref{fig:NLOreal} contribute. Before performing the $q_r$-derivative operations and the limit $q_r\to 0$ as in Eq.~(\ref{eq:QQFFQQA}), the amplitude of Feynman diagram in Fig.~\ref{fig:NLOreal}(a) could be written in a general form as $F(z,q_r)G(b, b', b'')\delta(1-z-\zeta_1+2q_r^+/p^+)$, where $G(b, b', b'')$ represents the color structure with color indices $b,b',b''$ listed in the figure, and $F(z,q_r)$ denotes the rest of the amplitude.
Then the amplitude of diagram in Fig.~\ref{fig:NLOreal}(b) could be obtained from that in Fig.~\ref{fig:NLOreal}(a) by performing  charge conjugation, as well as the replacements $\zeta_1 \to -\zeta_1$ and $q_r\to - q_r$, which give $(-1)^{S+\delta_{s,a}}F(z,-q_r)G^\dagger(b,b',b'')\delta(1-z+\zeta_1-2q_r^+/p^+)$. Therefore, the addition of these two diagrams is given by
\begin{eqnarray}
I_{a+b}
&=& F(z,q_r)G(b,b',b'')\delta\left(1-z-\zeta_1+\frac{2q_r^+}{p^+}\right)
\nonumber \\
&+ &
(-1)^{S+\delta_{s,a}}\,
F(z,-q_r)G^\dagger(b,b',b'')\delta\left(1-z+\zeta_1-\frac{2q_r^+}{p^+}\right) .
\label{eq:iab}
\end{eqnarray}
For producing a $S$-wave final-state $\cc$-pair, we can set the relative momentum $q_r$ to zero
and find
\begin{align}
I_{a+b}^S=
F(z,0)\left[G(b,b',b'')\delta(1-z-\zeta_1)+(-1)^{S+\delta_{s,a}}G^\dagger(b,b',b'')\delta(1-z+\zeta_1)\right] ,
\end{align}
which has two general structures depending on the color indices $b$ and $b'$.
If only one of the two indices is color octet, we have $G(b,b',b'')=G^\dagger(b,b',b'')$.
Multiplied with the complex conjugate of the amplitude, we obtain the first type of
the $\delta$-function structure:
\begin{align}\label{eq:DPMOne}
\left[\delta(1-z-\zeta_1)+(-1)^{S+\delta_{s,a}}\delta(1-z+\zeta_1)\right] \left[\delta(1-z-\zeta_2)+(-1)^{S+\delta_{s,a}}\delta(1-z+\zeta_2)\right].
\end{align}
If both $b$ and $b'$ are color octet, we have $G(b,b',b'')=\text{Tr}[t^{(F)}_b t^{(F)}_{b'} t^{(F)}_{b''}]$
with $t^{(F)}_b$ the generator of fundamental representation of SU(3) color.
Multiplied with the complex conjugate of the amplitude, we obtain the second type of the $\delta$-function structure:
%==========================
\begin{linenomath*}
\begin{align}
\begin{split}\label{eq:DPMEight}
&(N_c^2-2)\left[\DAA\DXX+\DBB\DYY\right]\\
&\hspace{1cm}
- (-1)^{S+\delta_{s,a}} 2\left[\DAA\DYY+\DBB\DXX\right].
\end{split}
\end{align}
\end{linenomath*}
%==========================
The $\delta$-function structures in Eqs.~(\ref{eq:DPMOne}) and (\ref{eq:DPMEight}) exhaust
all possible $\delta$-function structures of the NLO FFs to a $S$-wave $\cc$-pair.
Our explicit calculations in Ref.~\cite{\Swave} confirm the conclusion of above analysis.

For producing a $P$-wave final state $\cc$-pair, we need to take the $q_r$-derivative before setting $q_r$ to zero. From Eq.~(\ref{eq:iab}), we find the amplitude is a linear combination of
%==========================
\begin{align}
&F'(z,0)\left[G(b,b',b'')\delta(1-z-\zeta_1)+(-1)^{L+S+\delta_{s,a}}G^\dagger(b,b',b'')\delta(1-z+\zeta_1)\right],
\end{align}
and
\begin{align}
& F(z,0)\left[G(b,b',b'')\delta'(1-z-\zeta_1)+(-1)^{L+S+\delta_{s,a}}G^\dagger(b,b',b'')\delta'(1-z+\zeta_1)\right],
\end{align}
%==========================
where we have replaced $-(-1)^{S+\delta_{s,a}}$ by $(-1)^{L+S+\delta_{s,a}}$ since $L=1$.
Similar to the $S$-wave case, multiplying the above amplitude with its complex conjugate,
we obtain three $\delta$-function structures for each color combination.
The definitions of these structures are given in Appendix~\ref{app:results}.
The example calculations in next two sections, and our full results in Appendix~\ref{app:results}
clearly confirm the conclusions of our general analysis based on the symmetries.

%%%%%%%%%%%%%%%%%%%%%%%%%%%%%%%%%%%%%%%%%%%%%%%%%
\section{LO coefficients}\label{sec:LO}

%%%%%%%%%%%%%%%%%%%
\begin{figure}
\begin{center}
\includegraphics[width=0.6\textwidth]{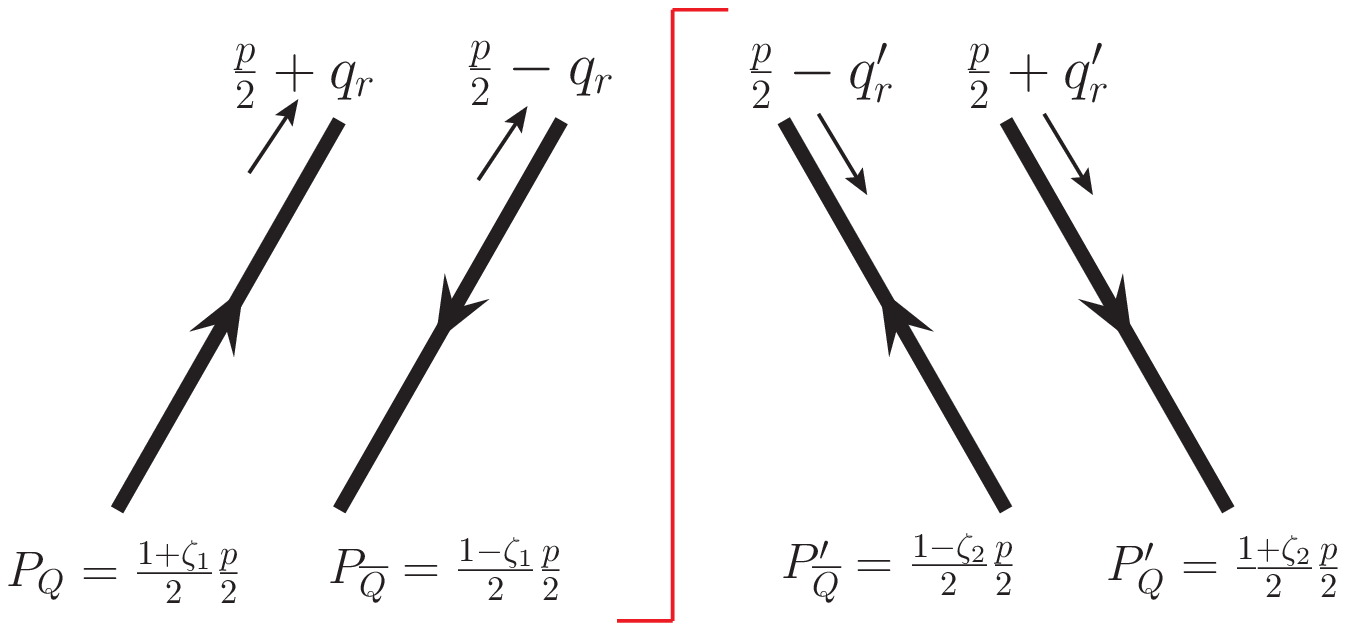}
\caption{Cut-diagram representation of ${\cal D}_{[\cc(s^{[b]})]\to [\cc(i^{[b']})]}$ at leading order.
\label{fig:LO}}
\end{center}
\end{figure}
%%%%%%%%%%%%%%%%%%%

A general cut-diagram representation for ${\cal D}^\text{LO}_{[\cc(s^{[b]})]\to \cc(i^{[b']})}$ is shown in Fig. (\ref{fig:LO}), with all momenta labeled explicitly.
At this order, the LDME in Eq.~(\ref{eq:QQFFNR2}) is proportional to $\delta_{n,n'}$, which leads to
%==========================
\begin{linenomath*}
\begin{align}\label{eq:QQFFNRmatchLO}
{\cal D}^{\text{LO}}_{[\cc(s^{[b]})]\to [\cc(i^{[b']})]}(z,\zeta_1,\zeta_2;m_Q,\mu_0) = \hat{d}^{\text{ LO}}_{[\cc(s^{[b]})]\to [\cc(i^{[b']})]}(z,\zeta_1,\zeta_2;m_Q,\mu_0).
\end{align}
\end{linenomath*}
%==========================
For our purpose of producing a $P$-wave non-relativistic $\cc$-pair, Eqs.~(\ref{eq:QQFFQQ}) and (\ref{eq:QQFFQQA})
could be reduced to,
%==========================
\begin{linenomath*}
\begin{align}
&{\cal D}^{\text{LO}}_{[\cc(s^{[b]})]\to [\cc(i^{[b']})]}(z,\zeta_1,\zeta_2;m_Q,\mu_0)
=
\frac{\delta(1-z)}{N_s N_b N_{i}^\text{NR} N_{b'}^\text{NR}}
{\cal M}^{\text{LO}}_{[\cc(s^{[b]})]\to [\cc(i^{[b']})]}(p, z, \zeta_1, \zeta_2)\, ,\label{eq:QQFFQQLO}
\end{align}
\end{linenomath*}
and
\begin{linenomath*}
\begin{align}
&{\cal A}^{\text{LO}}_{[\cc(s^{[b]})]\to [\cc(i^{[b']})]}(p,z,\zeta_1)
=
\lim_{q_r\to0}
\frac{\text{d}}{\text{d}{q_r^{\alpha}}}
\Big\{
2\ \delta(\zeta_1-\frac{2q_r^+}{p_c^+})
\nonumber
\\
&
\hspace{5cm}
\times
{\cal \bar{A}}^{\text{LO}}_{[\cc(s^{[b]})]\to [\cc(i^{[b']})]}(q_1=q_r)
\Gamma_{i}^{\text{NR}}(p)\, C_{b'}^\text{NR}
\Big\},
\label{eq:QQFFQQALO}
\end{align}
\end{linenomath*}
%==========================
respectively.  The above three equations, plus Eq.~(\ref{eq:QQFFQQM2}),
form the basis of our calculation for $\hat{d}^{\text{ LO}}_{[\cc(s^{[b]})]\to [\cc(i^{[b']})]}(z,\zeta_1,\zeta_2;m_Q,\mu_0)$.
Based on the discussion in Sec.~\ref{sec:color}, we find that only independent FFs at LO are those with both initial and final $\cc$-pair being color singlet. In the following, we show the detailed calculations of two examples, while presenting all nonzero results in Appendix \ref{sec:LOresults}.

First, we consider LO contribution to fragmentation process:\ $[\cc(\vone)] \to [\cc(\CScPj)]$.
From Eqs.~(\ref{eq:QQFFQQM2}) and (\ref{eq:QQFFQQALO}), we have
%==========================
\begin{linenomath*}
\begin{align}\label{eq:QQFFQQLOPTr1}
\begin{split}
&{\text{Tr}}\left[\Gamma_v(p_c)\, C_1\, {\cal A}^{\text{LO}}_{[\cc(\vone)]\to [\cc(\CScPj)]}(p, z, \zeta_1)\right]
\\&
\hspace{2cm}
=
\lim_{q_r\to0}
\frac{\text{d}}{\text{d}{q_r^{\alpha}}}
\int\frac{d^D q_1}{(2\pi)^D}\delta^D(q_1-q_r)
\bigg\{ 2\,\delta(\zeta_1-\frac{2 q_1^+}{p^+})\,
\text{Tr}_c\left[\frac{1}{\sqrt{N_c}}\frac{1}{\sqrt{N_c}}\right]
\\&
\hspace{2cm}
\times
\text{Tr}_{\gamma}\left[\frac{\gamma\cdot {\hat{n}}}{4 p\cdot {\hat{n}}}\frac{1}{\sqrt{8m_Q^3}}
\left(\frac{\slashed{p}}{2}-\slashed{q}_r-m_Q\right)
\gamma^\beta\left(\frac{\slashed{p}}{2}+\slashed{q}_r+m_Q\right)\right]
\bigg\}
\\
&\hspace{2cm}
=\frac{\delta'(\zeta_1)}{\sqrt{2m_Q^3}(p^+)^2}n^\alpha(4\, m_Q^2 n^\beta-p^+ p^\beta),
\end{split}
\end{align}
\end{linenomath*}
%==========================
where $``\text{Tr}_c"$ ($``\text{Tr}_\gamma"$) denotes the trace of color matrices ($\gamma$-matrices).  In deriving Eq.~(\ref{eq:QQFFQQLOPTr1}),
we used the projection operators defined in Appendix~A of Ref.~\cite{\Swave} as well as the fact $p_c=p$.
Substituting our result in Eq.~(\ref{eq:QQFFQQLOPTr1}) into Eq.~(\ref{eq:QQFFQQM2}), and then Eq.~(\ref{eq:QQFFQQLO}), and using Eq.~(\ref{eq:QQFFNRmatchLO}), we obtain
%==========================
\begin{linenomath*}
\begin{subequations}
\begin{align}
\hat{d}^{\text{ LO}}_{[\cc(\vone)]\to [\cc(\CScPz)]}(z,\zeta_1,\zeta_2;m_Q,\mu_0)
&=\frac{1}{D-1}\frac{1}{2m_Q^3}\,\delta(1-z)\,\delta'(\zeta_1)\,\delta'(\zeta_2),\\
\hat{d}^{\text{ LO}}_{[\cc(\vone)]\to [\cc(\CScPa)]}(z,\zeta_1,\zeta_2;m_Q,\mu_0)
&=0,\\
\hat{d}^{\text{ LO}}_{[\cc(\vone)]\to [\cc(\CScPb)]}(z,\zeta_1,\zeta_2;m_Q,\mu_0)
&=\frac{1}{(D-1)(D+1)}\frac{1}{m_Q^3}\,\delta(1-z)\,\delta'(\zeta_1)\,\delta'(\zeta_2),
\end{align}\label{eq:ff-p-lo}
\end{subequations}
\end{linenomath*}
%==========================

\noindent
where $D=4-2\epsilon$.
The zero result in the second equation is expected from Lorentz invariance, as explained in Sec.~\ref{sec:lorentz}. The fact that the right-hand-side of the first and the third equations in
Eq.~(\ref{eq:ff-p-lo}) are odd in both $\zeta_1$ and $\zeta_2$ is consistent with our analysis
above the equation~\eqref{eq:parity23}.

For the second example, we consider LO contribution to fragmentation process:\
$[\cc(\aone)]\to [\cc(\CScPa)]$.  Similar to Eq.~(\ref{eq:QQFFQQLOPTr1}),
the corresponding trace is
%==========================
\begin{linenomath*}
\begin{align}\label{eq:QQFFQQLOPTr2}
\begin{split}
&{\text{Tr}}\left[\Gamma_a(p_c)\, C_1\, {\cal A}^{\text{LO}}_{[\cc(\aone)]\to [\cc(\CScPa)]}(p, z, \zeta_1)\right]
=
\lim_{q_r\to0}
\frac{\text{d}}{\text{d}{q_r^{\alpha}}}
\bigg\{2\, \delta(\zeta_1-\frac{2 q_r^+}{p^+})\,
\text{Tr}_c\left[\frac{1}{\sqrt{N_c}}\frac{1}{\sqrt{N_c}}\right]
\\&
\hspace{2cm}
\times
\text{Tr}\left[\frac{\gamma\cdot {\hat{n}}\, \gamma_5-\gamma_5\, \gamma\cdot {\hat{n}}}{8 p\cdot {\hat{n}}}\frac{1}{\sqrt{8m_Q^3}}
\left(\frac{\slashed{p}}{2}-\slashed{q}_r-m_Q\right)
\gamma^\beta\left(\frac{\slashed{p}}{2}+\slashed{q}_r+m_Q\right)\right]
\bigg\}
\\
&\hspace{2cm}
=\frac{\delta(\zeta_1)}{4p^+\sqrt{2m_Q^3}} \text{Tr}\left[\gamma_5\,\gamma^\alpha\,\gamma^\beta\,\slashed{\hat{n}}\,\slashed{p}\right].
\end{split}
\end{align}
\end{linenomath*}
%==========================
The Lorentz structure is exactly the same with our analysis in Sec.~\ref{sec:lorentz}.
Finally, by substituting our result in Eq.~(\ref{eq:QQFFQQLOPTr2}) into Eq.~(\ref{eq:QQFFQQM2}), and then Eq.~(\ref{eq:QQFFQQLO}), and using Eq.~(\ref{eq:QQFFNRmatchLO}), we obtain
%==========================
\begin{linenomath*}
\begin{align}
\hat{d}^{\text{ LO}}_{[\cc(\aone)]\to [\cc(\CScPa)]}(z,\zeta_1,\zeta_2;m_Q)
=\frac{D-3}{m_Q^3 (D-1)}\delta(\zeta_1)\delta(\zeta_2)\delta(1-z)
.
\end{align}
\end{linenomath*}
%==========================
As expected from our discussion above the equation~\eqref{eq:parity23}, this result is even in both $\zeta_1$ and $\zeta_2$.

%%%%%%%%%%%%%%%%%%%%%%%%%%%%%%%%%
%                                                   						  %
%                                   NLOcalculation					  %
%  											   	  %
%%%%%%%%%%%%%%%%%%%%%%%%%%%%%%%%%
\section{NLO coefficients} \label{sec:NLO}

\begin{figure}[htb]
\begin{center}
\includegraphics[width=0.45\textwidth]{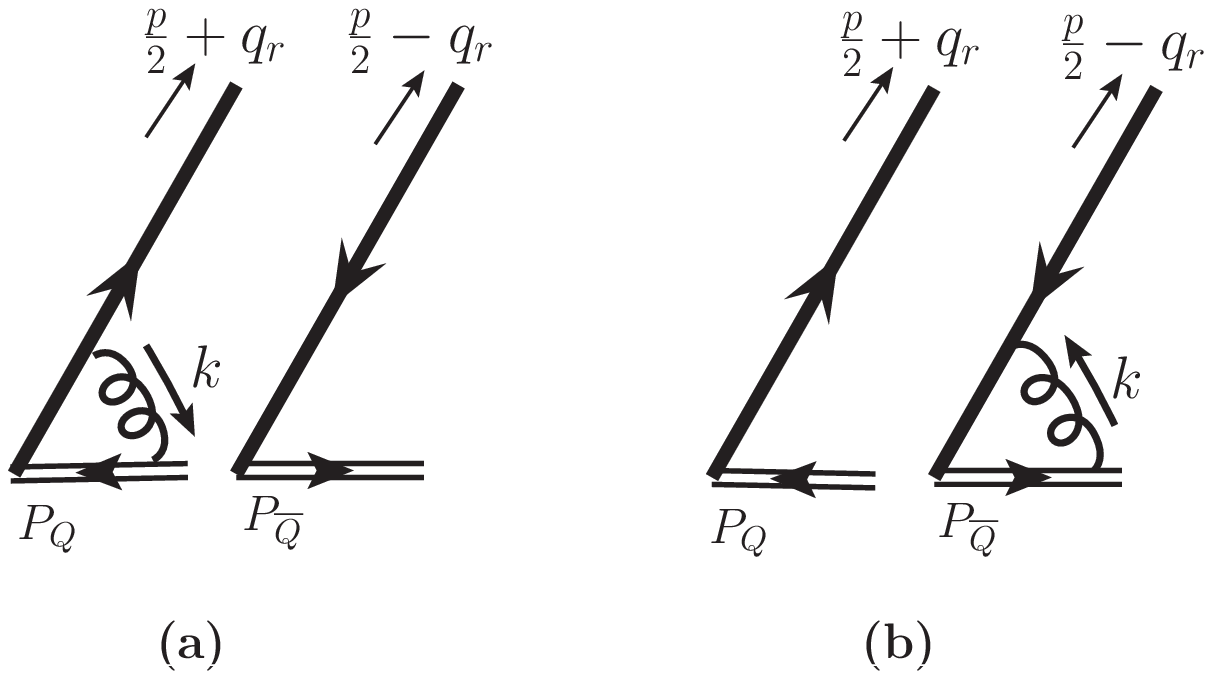}\hspace{0.5cm}
\includegraphics[width=0.43\textwidth]{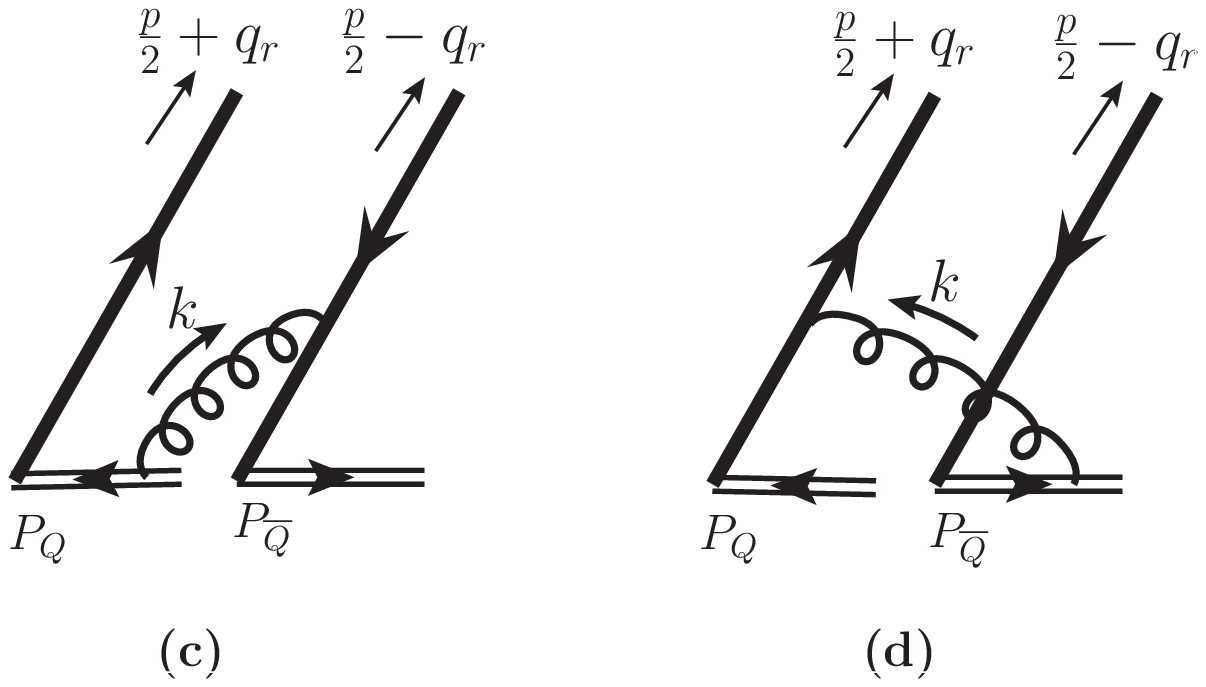}\\
\includegraphics[width=0.19\textwidth]{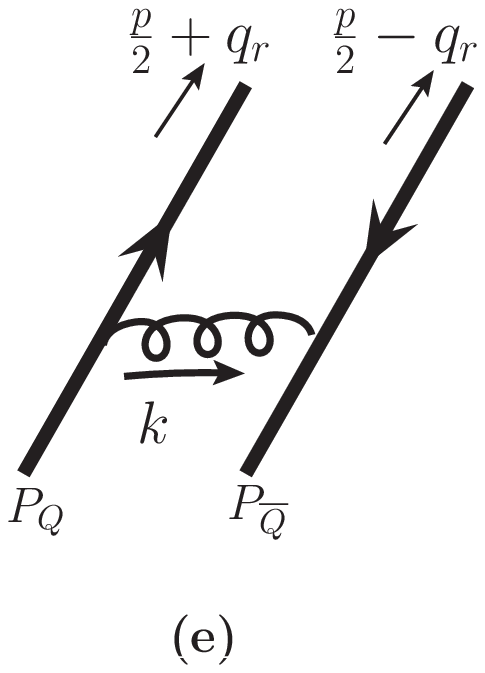}\hspace{0.5cm}
\includegraphics[width=0.7\textwidth]{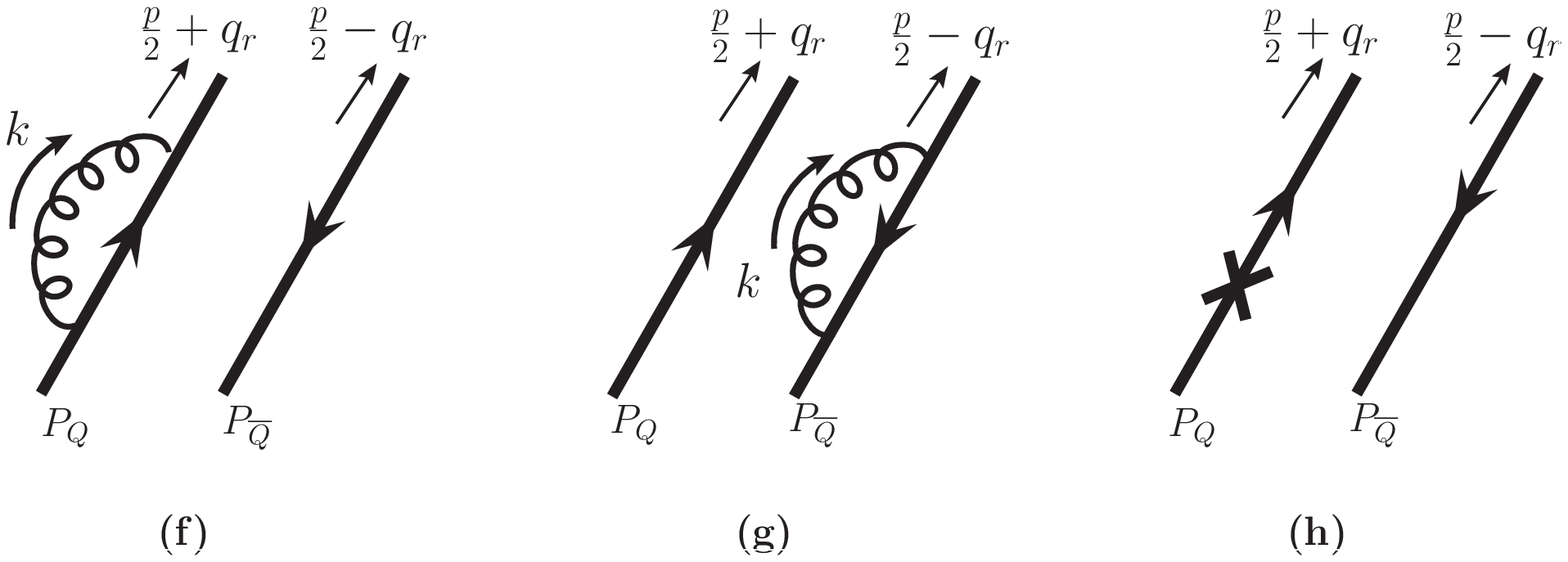}\\
\includegraphics[width=0.19\textwidth]{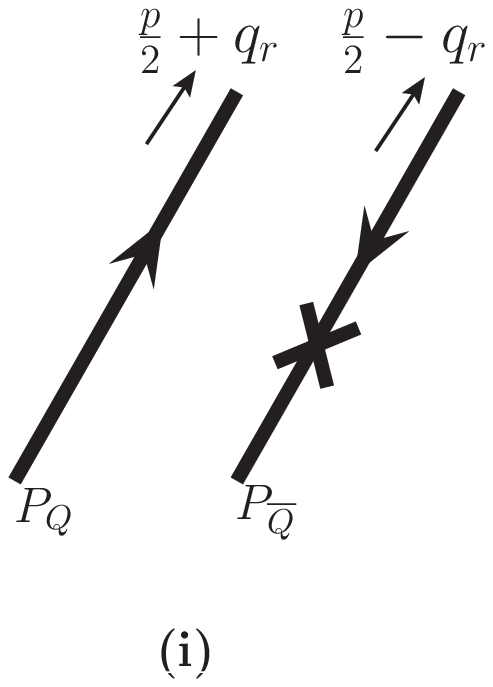}\hspace{0.9cm}
\includegraphics[width=0.18\textwidth]{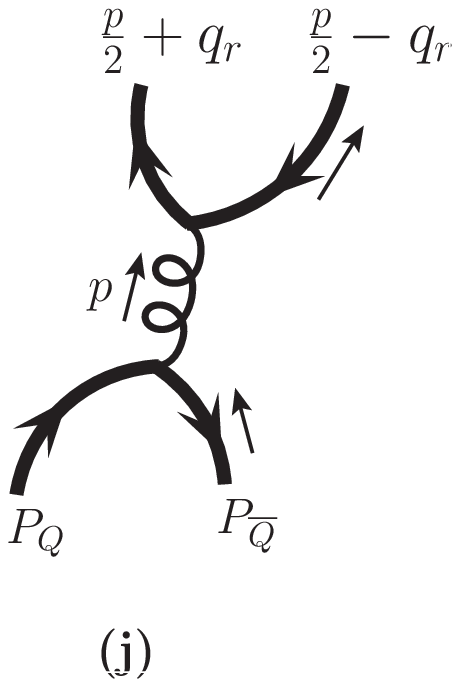}\hspace{0.9cm}
\includegraphics[width=0.42\textwidth]{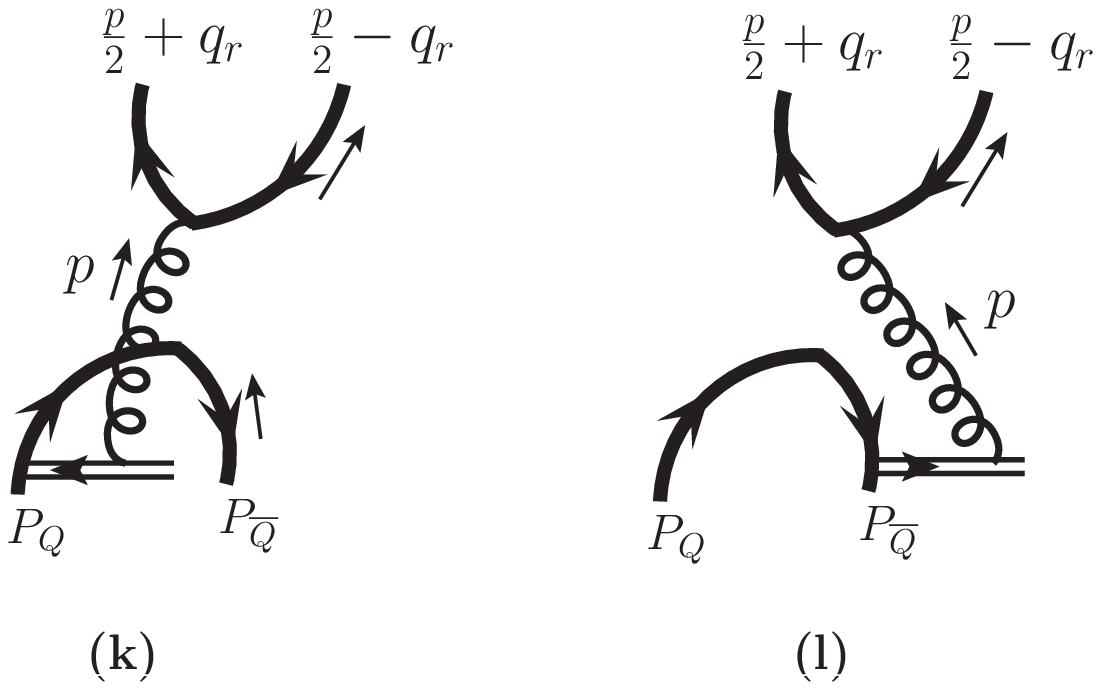}\\
\caption{Feynman diagrams for virtual correction at NLO.
\label{fig:NLOvirtual}}
\end{center}
\end{figure}

\begin{figure}
\begin{center}
\includegraphics[width=0.45\textwidth]{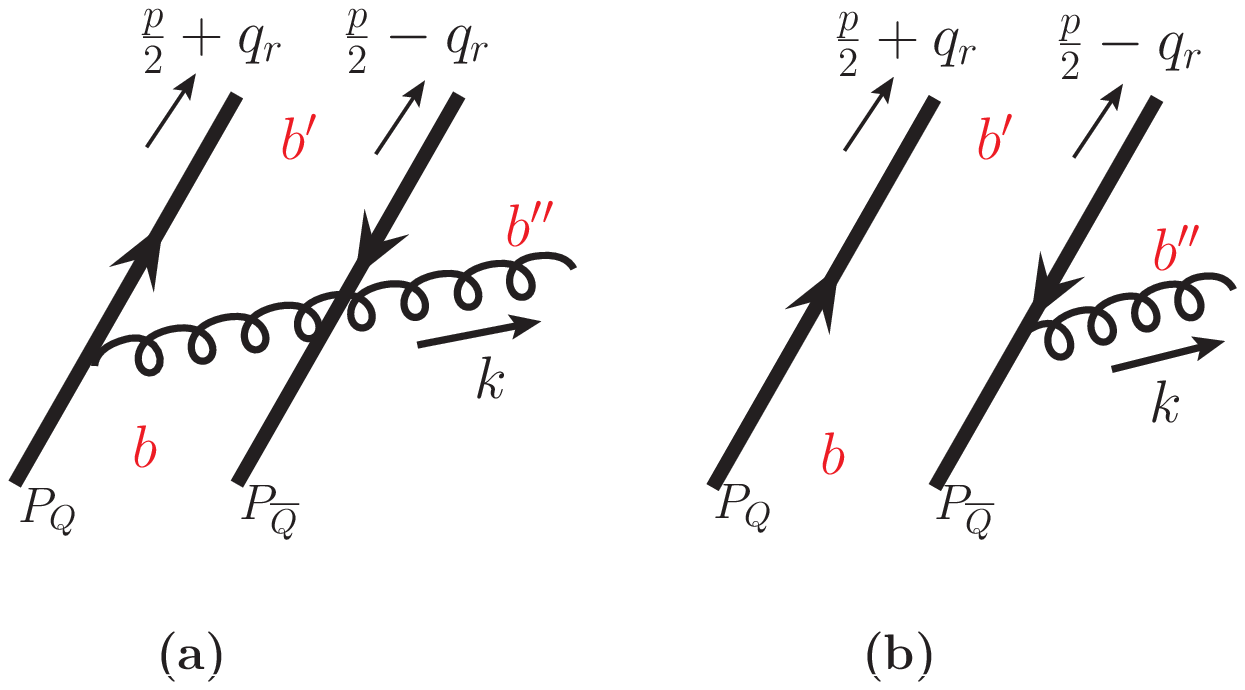}
\hspace{0.9cm}
\includegraphics[width=0.45\textwidth]{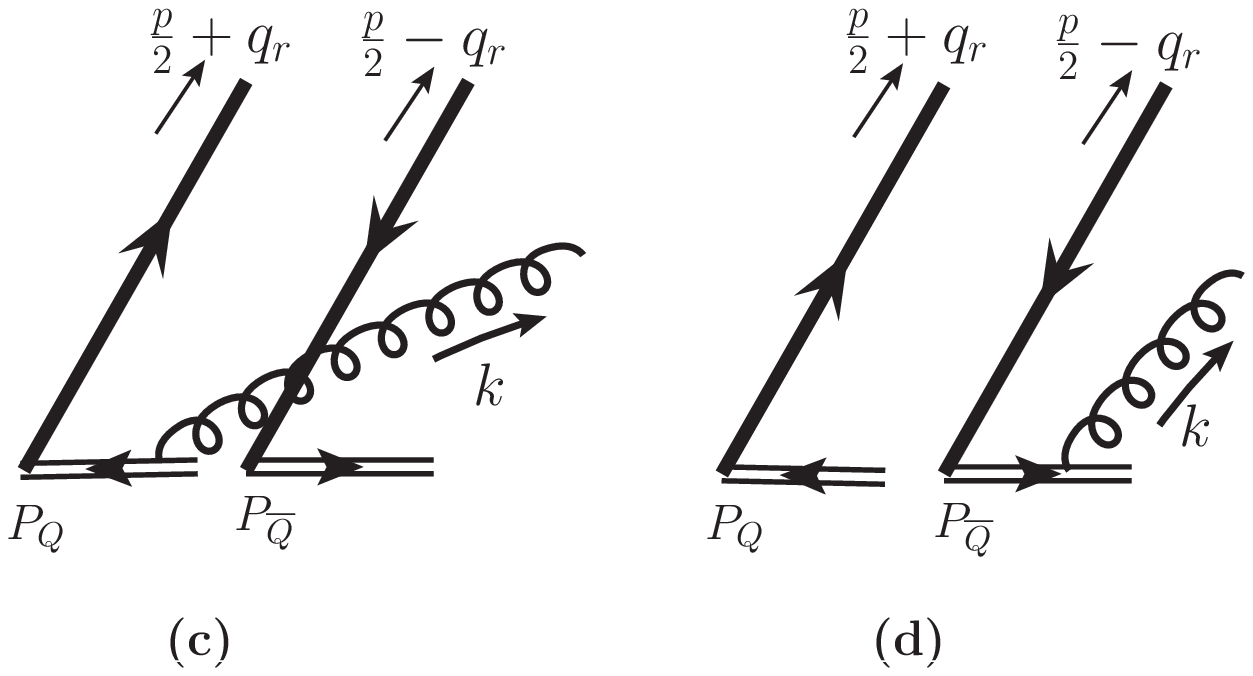}
\caption{Feynman diagrams for real correction at NLO.
\label{fig:NLOreal}}
\end{center}
\end{figure}

In this section, we calculate the short-distance coefficients in Eq.~(\ref{eq:QQFFNR2}) at NLO in $\as$.
We first expand both sides of Eq.~(\ref{eq:QQFFNR2}) to NLO,
%==========================
\begin{linenomath*}
\begin{align}\label{eq:QQFFNRmatchNLO}
\begin{split}
{\cal D}^{\text{NLO}}_{[\cc(\kappa)]\to [\cc(n')]}(z,\zeta_1,\zeta_2;m_Q,\mu_0) &= \hat{d}^{\text{ NLO}}_{[\cc(\kappa)]\to [\cc(n')]}(z,\zeta_1,\zeta_2;m_Q,\mu_0,\mu_\Lambda)
\\
&\hspace{-1cm}
+
\sum_{[\cc(n)]} \hat{d}^{\text{ LO}}_{[\cc(\kappa)]\to [\cc(n)]}(z,\zeta_1,\zeta_2;m_Q,\mu_0){{\langle{\cal O}_{\cc[n]}^{\cc[n']}(\mu_\Lambda)\rangle^{\text{NLO}}}},
\end{split}
\end{align}
\end{linenomath*}
%==========================
Generally, the LHS of Eq.~(\ref{eq:QQFFNRmatchNLO}) has both virtual and real contributions,
which are represented
by Feynman diagrams in Figs.~\ref{fig:NLOvirtual} and \ref{fig:NLOreal}, respectively.
For any specific subprocess, we could use symmetry constraints derived in Sec.~\ref{sec:lorentz}
to simplify our calculations.

%==========================
%==========================

%==========================
%==========================

We use dimensional regularization to regularize all kinds of divergences, including
the ultra-violet (UV) divergence, infrared (IR) divergence, rapidity divergence, and Coulomb divergence.
We apply the technical trick discussed below Eq.~(\ref{eq:QQFFQQA}) to our calculations.
Consequently, the Coulomb divergence does not appear in our derivations.
As shown in Appendix~\ref{app:coulomb}, the Coulomb divergence in our calculations
are effectively absorbed into the NRQCD LDMEs.
The rapidity divergence comes from region $k\cdot \hat{n}\to 0$,
with $k$ the momentum of radiated gluon and $\hat{n}$ the light-cone vector.
This region overlaps with UV region and could lead to a double pole.
By adding up all diagrams, the rapidity divergence cancels.

After summing over all diagrams in Figs.~\ref{fig:NLOvirtual} and \ref{fig:NLOreal},
there are still leftover UV and IR divergences in ${\cal D}^{\text{NLO}}_{[\cc(\kappa)]\to [\cc(n')]}(z,\zeta_1,\zeta_2;m_Q)$
for producing a $P$-wave NRQCD $\cc$-pair with the spin-color state $[\cc(n')]$.
The UV divergence is cancelled by pQCD renormalization of the operator defining the FFs
in the same manner as the calculation of FFs to produce $S$-wave $\cc$-pairs \cite{\Swave}.
If NRQCD factorization is valid to NLO,
the leftover IR divergence should be exactly the same as the IR divergence in the NLO LDMEs
on the right hand side of Eq.~(\ref{eq:QQFFNRmatchNLO}), leaving NLO short-distance coefficient IR safe.

In this paper, we calculated all short-distance coefficients of FFs through a $P$-wave NRQCD $\cc$-pair
up to NLO in $\as$.  From our explicit calculations, we find that the leftover IR divergence cancels
for all fragmentation channels at this order.

In the rest of this section, we present the calculation of the short-distance coefficient
for the fragmentation channel:\ $[\cc(\aone)]\to [\cc(\COaPa)]$ to demonstrate the
cancellation of infrared divergence in NRQCD factorization formalism.
From Eq.~(\ref{eq:QQFFNRmatchNLO}), the NLO short-distance coefficient for this channel
is given by
%==========================
\begin{linenomath*}
\begin{align}\label{eq:QQFFNRmatchNLOP}
\begin{split}
\hat{d}^{\text{ NLO}}_{[\cc(\aone)]\to [\cc(\COaPa)]}(z,\zeta_1,\zeta_2;m_Q,\mu_0,\mu_\Lambda)
&=
{\cal D}^{\text{NLO}}_{[\cc(\aone)]\to [\cc(\COaPa)]}(z,\zeta_1,\zeta_2;m_Q,\mu_0)
\\
&\hspace{-2cm}
-
 \hat{d}^{\text{ LO}}_{[\cc(\aone)]\to [\cc(\CSaSz)]}(z,\zeta_1,\zeta_2;m_Q,\mu_0)
 {{\Big\langle{\cal O}_{[\cc(\CSaSz)]}^{[\cc(\COaPa)]}(\mu_\Lambda)\Big\rangle^{\text{NLO}}}},
\end{split}
\end{align}
\end{linenomath*}
%==========================
For this channel, calculation of
${\cal D}^{\text{NLO}}_{[\cc(\aone)]\to [\cc(\COaPa)]}(z,\zeta_1,\zeta_2;m_Q,\mu_0)$
involves only real correction from Feynman diagrams shown in Fig.~\ref{fig:NLOreal}.
The calculation of these diagrams is complicated by both UV and IR divergences.
We remove the UV divergence by the UV counter-term in $\overline{\text{MS}}$ scheme, which is associated with pQCD renormalization of the operator defining $[\cc(\aone)]$.  This is just like what we did in our calculation of $\cc$ FFs to $S$-wave $\cc$-pairs \cite{\Swave}.  After this $\overline{\text{MS}}$ renormalization procedure, the calculated fragmentation function ${\cal D}^{\text{NLO}}_{[\cc(\aone)]\to [\cc(\COaPa)]}(z,\zeta_1,\zeta_2;m_Q,\mu_0)$ has only IR divergence left,
%==========================
\begin{linenomath*}
\begin{align}\label{eq:NLOP1}
\begin{split}
{\cal D}^{\text{NLO}}_{[\cc(\aone)]\to [\cc(\COaPa)]}
=&
\frac{\alpha_s z \,C_F}{24\pi\,m_Q^3(N_c^2-1)}
\left\{ (1-z)\DPlusOne+\DPlusPOne+\frac{1}{(1-z)}\DPlusPPOne
\right\}
\Big(\text{ln}\big[\frac{\mu_0^2}{4m_Q^2}\big]-\frac{2}{3}\Big)
\\
&\hspace{-1cm}
+
\frac{\alpha_s \,C_F}{18\pi\,m_Q^3(N_c^2-1)}
\left(\frac{\pi\mu^2}{m_Q^2}\right)^\epsilon\Gamma(1+\epsilon)
\left\{
\left(-\frac{3}{\epsilon_{\text{IR}}}+1\right)\DeltaZero\,\delta(1-z)
\right.
\\
&\hspace{-1cm}
-\frac{\DPlusOne}{4}\left[-6\left(\frac{1}{1-z}\right)_+ +5z^2+7z-6(z-1)z\,\text{ln}(1-z)+6\right]
\\
&\hspace{-1cm}
\left.
+
\frac{\DPlusPOne}{4}\frac{z}{(1-z)}[z+6(z-1)\text{ln}(1-z)-4]
+
\frac{\DPlusPPOne}{4}\frac{z}{(1-z)}[-6\,\text{ln}(1-z)-1]
\right\},
\end{split}
\end{align}
\end{linenomath*}
%==========================
where $\mu$ is the renormalization scale, $\mu_0$ is the pQCD factorization scale for input FFs, and the $2/3$ along with $\mylog{\mu_0^2/4m_Q^2}$ came from the $\epsilon$-dependence of $\hat{d}_{[\cc(\aeight)]\to[\cc(\COaPa)]}^{\text{ LO}}$, which is a part of the UV counter-term.  As expected from the general symmetry analysis in Sec.~\ref{sec:charge}, and
as shown in Eq.~(\ref{eq:NLOP1}), only possible structures of $\zeta_1$ and $\zeta_2$ dependence
for ${\cal D}^{\text{NLO}}_{[\cc(\aone)]\to [\cc(\COaPa)]}(z,\zeta_1,\zeta_2;m_Q,\mu_0)$
are given by those $\Delta$-functions: $\DPlusOne$, $\DPlusPOne$ and $\DPlusPPOne$,
defined in Appendix~\ref{app:results}.

Note that since both $\DPlusPOne$ and $\DPlusPPOne$ vanish at $z\to1$, the denominator $(1-z)$
in Eq.~(\ref{eq:NLOP1}), so as that in Eq.~(\ref{eq:dNLOP1}) below, does not exhibit a pole as $z\to 1$.  However, the partonic fragmentation function in Eq.~(\ref{eq:NLOP1}) still shows IR divergence, as indicated by $1/\epsilon_{\rm IR}$, which is expected to be cancelled by the second term on the RHS of Eq.~(\ref{eq:QQFFNRmatchNLOP}).

For the second term on the right-hand-side of the Eq.~(\ref{eq:QQFFNRmatchNLOP}), the LDME has been calculated in previous publications~\cite{Bodwin:1994jh,Huang:1996fa}
%==========================
\begin{linenomath*}
\begin{align}\label{eq:LDMENLO}
\begin{split}
 {{\Big\langle{\cal O}_{[\cc(\CSaSz)]}^{[\cc(\CSaPa)]}(\mu_\Lambda)\Big\rangle^{\text{NLO}}}}
 =
 \frac{2\alpha_s}{3\,\pi \,m_Q^2 N_c}(4\pi e^{-\gamma_E})^\epsilon\left(\frac{\mu^2}{\mu_\Lambda^2}\right)^\epsilon\frac{1}{\epsilon_{\text{IR}}},
\end{split}
\end{align}
\end{linenomath*}
%==========================
where we have chosen $\overline{\rm MS}$ renormalization scheme, and $\mu$ ($\mu_\Lambda$) is the renormalization (NRQCD factorization) scale.
The LO short-distance coefficient  in Eq.~(\ref{eq:QQFFNRmatchNLOP}) has been calculated in our previous paper as \cite{\Swave},
%==========================
\begin{linenomath*}
\begin{align}\label{eq:dhatLO}
\begin{split}
 \hat{d}^{\text{ LO}}_{[\cc(\aone)]\to \cc[(\CSaSz)]}(z,\zeta_1,\zeta_2;m_Q)
 =
 \frac{1}{2\,m_Q}\delta(1-z)\delta(\zeta_1)\delta(\zeta_2).
\end{split}
\end{align}
\end{linenomath*}
%==========================
Therefore the second term of Eq.~(\ref{eq:QQFFNRmatchNLOP}) is
%==========================
\begin{linenomath*}
\begin{align}\label{eq:2ndterm}
\begin{split}
&
 \hat{d}^{\text{ LO}}_{[\cc(\aone)]\to[\cc(\CSaSz)]}(z,\zeta_1,\zeta_2;m_Q) {{\Big\langle{\cal O}_{[\cc(\CSaSz)]}^{[\cc(\CSaPa)]}(\mu_\Lambda)\Big\rangle^{\text{NLO}}}}
 \\
 &\hspace{2cm}
 =
 \frac{\alpha_s}{4\,\pi \,m_Q^3 N_c}(4\pi e^{-\gamma_E})^\epsilon\left(\frac{\mu^2}{\mu_\Lambda^2}\right)^\epsilon\frac{1}{\epsilon_{\text{IR}}}\delta(1-z)\delta(\zeta_1)\delta(\zeta_2).
\end{split}
\end{align}
\end{linenomath*}
%==========================
Substitute Eqs.~(\ref{eq:NLOP1}) and (\ref{eq:2ndterm}) into Eq.~(\ref{eq:QQFFNRmatchNLOP}),
we find that the IR divergence is cancelled exactly between the first and second terms in
Eq.~(\ref{eq:QQFFNRmatchNLOP}), and the finite remainder after the IR cancelation
is effectively the NLO short-distance coefficient,
%==========================
\begin{linenomath*}
\begin{align}\label{eq:dNLOP1}
\begin{split}
&
\hat{d}^{\text{ NLO}}_{[\cc(\aone)]\to \cc[(\COaPa)]}(z,\zeta_1,\zeta_2;m_Q,\mu_0,\mu_\Lambda)
=\frac{\as z}{24\pi\, m_Q^3 N_c}
\Big\{
2\,\DeltaZero\,\delta(1-z)(-\LogIR + 2\logtwo + \frac{1}{3})
\\
&
\hspace{2cm}
+\big[\frac{\DPlusPPOne}{(1-z)} +\DPlusPOne +\DPlusOne(1-z)\big] (\frac{1}{2}\LogUV-\frac{1}{3})
\\
&
\hspace{2cm}
-\frac{\DPlusPPOne}{(1-z)}  R_{1}(z)- \frac{\DPlusPOne}{(1-z)} R_{2}(z)-\DPlusOne \,R_{3}(z)
\Big\}
,
\end{split}
\end{align}
\end{linenomath*}
%==========================
The $R$-functions in Eq.~(\ref{eq:dNLOP1}) are defined as
%==========================
\begin{linenomath*}
\begin{align}
%=========R a1==========
\begin{split}
&
R_{1}(z)
=
\mylog{2-2z}+\frac{1}{6}
,
\end{split}\\
%=========R a2==========
\begin{split}
&
R_{2}(z)
=
(1-z)\mylog{2-2z}-\frac{1}{6}z+\frac{2}{3}
,
\end{split}\\
%=========R a3==========
\begin{split}
&
R_{3}(z)
=
-\frac{1}{(1-z)_+}+(1-z)\mylog{2-2z}+\frac{5}{6}z+\frac{7}{6}.
\end{split}
\end{align}
\end{linenomath*}
%==========================
A complete list of our results are given in Appendix~\ref{app:results}.

%%%%%%%%%%%%%%%%%%%%%%%%%%%%%%%%%
%                                                   						  %
%                                   summary						  %
%  											   	  %
%%%%%%%%%%%%%%%%%%%%%%%%%%%%%%%%%
\section{summary and conclusions} \label{sec:summary}

We calculated the heavy quarkonium FFs at the input scale $\mu_0 \gtrsim 2m_Q$ in terms of the NRQCD factorization formalism.  We evaluated all short-distance coefficients for a perturbatively produced heavy quark pair to evolve into a $P$-wave non-relativistic heavy quark pair to the first non-trivial order in $\as$.  Along with our companion paper \cite{\Swave}, in which we presented our evaluation of all perturbative coefficients for a produced heavy quark pair to evolve into a $S$-wave non-relativistic heavy quark pair, we effectively expressed all non-perturbative heavy quarkonium FFs (at least  {\it ten} unknown functions for each heavy quarkonium state produced) in terms of very few NRQCD LDMEs per quarkonium state with perturbatively calculated coefficients for their dependence on momentum fractions, $z$, $\zeta_1$ and $\zeta_2$.

Although there is no formal proof of NRQCD factorization approach to evaluate the heavy quarkonium FFs, we found that all infrared divergences of the FFs at this first non-trivial order are exactly the same as the NLO expansion of NRQCD LDMEs, which ensures that the calculated short-distance coefficients are infrared safe.
In addition, we found that due to the underline symmetries of QCD, in particular, the charge conjugation symmetry, the structure (or the dependence on the momentum fractions) of all short-distance coefficients/contributions to the FFs are very compact, with only a few distinctive structures.  Just like any perturbative calculation of short-distance coefficients in a factorization approach, there are factorization scheme dependence for the calculated coefficients at NLO and beyond.  In this paper and its companion one \cite{\Swave}, we used the dimensional regularization and $\overline{\rm MS}$ factorization scheme.  It is straightforward to convert our results into any other regularization and factorization schemes.

In principle, the heavy quarkonium FFs are nonperturbative and universal, and their functional forms at the input scale, $\mu_0\gtrsim 2m_Q$, should be extracted from the experimental data.  In practice, it is difficult to extract such a large number of fragmentation functions to test the pQCD factorization formalism precisely.  The predictive power of the factorization formalism does rely on our knowledge of these input FFs.  With the heavy quark mass and the NRQCD factorization approach, we effectively expressed all these unknown FFs in terms of a few universal NRQCD matrix elements, and boosted the predictive power of the formalism.  We are aware of the fact that the NRQCD factorization approach to evaluate the heavy quarkonium FFs have not been proved to all orders in perturbative expansion of $\as$ and the pair's relative velocity, $v$.   With the cancelation of all IR divergences at the first non-trivial order from our explicit calculations, our effort in this and its companion paper is the first step to bridge the gap between the perturbative QCD factorization formalism and its phenomenological applications.

Since the heavy quarkonium FFs at the input scale are only quantities of the pQCD factorization formalism for heavy quarkonium production that are actually sensitive to the properties of individual heavy quarkonium state produced, understanding these input FFs in a more controlled way, other than simply fitting the data, should help us to gain valuable knowledge on how heavy quarkonia are actually emerged from produced heavy quark pairs.  Such formation of a bound quarkonium could take place in the vacuum, or in a high temperature and/or high density medium.  With heavy quark mass $m_Q \gg \Lambda_{\rm QCD}$, and the distinctive momentum scales of $m_Q v$, $m_Qv^2$, and etc., heavy quarkonium production in various environments could be a very important diagnosing tool in QCD condensed matter physics \cite{Brambilla:2010cs,Bodwin:2013nua}.

%=======================================================================
\section*{Acknowledgments}

We thank Zhong-Bo Kang and George Sterman for helpful discussions.
We also thank D. Yang and X. Wang for lots of communications 
for cross-checking the results in the color singlet channel.
This work was supported in part by the U. S. Department of Energy
under contract No. DE-AC02-98CH10886,
and the National Science Foundation under
grant Nos.~PHY-0354776, PHY-0354822 and PHY-0653342.

\appendix
%%%%%%%%%%%%%%%%%%%%%%%%%%%%%%%%%
%                                                   						  %
%                                   App I: app_coulomb				  %
%  											   	  %
%%%%%%%%%%%%%%%%%%%%%%%%%%%%%%%%%
\section{Coulomb divergence and the expansion of heavy quark relative momentum}\label{app:coulomb}

In this appendix, we justify our procedure of calculating the $P$-wave contributions by expanding the relative momentum of the final-state non-relativistic heavy quark pair, $q_r$, before the integration over the relative momentum of the initial-state perturbative heavy quark pair, $q_1$.

As shown in Eq.~\eqref{eq:QQFFQQA}, the general structure of one-loop amplitude of the $\cc$-pair FFs is of the following form,
\begin{linenomath*}
\begin{align}\label{eq:method1}
\begin{split}
A_1(\zeta_1)=\lim_{q_r\to0}\left(\prod_{j=0}^L
\frac{\text{d}}{\text{d}{q_r^{\alpha_j}}}\right)
\left\{
\int\frac{{\text d}^D q_1}{(2\pi)^D}\,
\delta(\zeta_1-\frac{2q_1^+}{p_c^+})\,\bar{A}(q_1,q_r)
\right\}
\, ,
\end{split}
\end{align}
\end{linenomath*}
where the $q_1$-integration should be performed before taking the derivatives with respect to $q_r$ and the limit $q_r\to 0$.  However, the calculation in this order is often very complicated. On the other hand, a similar calculation,
\begin{linenomath*}
\begin{align}\label{eq:method2}
\begin{split}
A_2(\zeta_1)=
\int\frac{{\text d}^D q_1}{(2\pi)^D}\,
\delta(\zeta_1-\frac{2q_1^+}{p_c^+}) \lim_{q_r\to0}\left(\prod_{j=0}^L
\frac{\text{d}}{\text{d}{q_r^{\alpha_j}}} \bar{A}(q_1,q_r)\right)
\, ,
\end{split}
\end{align}
\end{linenomath*}
could be carried out much easier due to the fact that the derivatives and the limit of $q_r$ were taken before performing the $q_1$-integration. In general, $A_1(\zeta_1)$ and $A_2(\zeta_1)$ are not necessary to be equal, unless the integration region of $q_1 \lesssim q_r \to 0$ is not important, which means that the integrand $\bar{A}(q_1,q_r)$ has no pole as $q_1 \to 0$ and $q_r \to 0$. Unfortunately, this condition is not satisfied by the process that we are considering here.

However, we show in this appendix that the difference between $A_1(\zeta_1)$ and $A_2(\zeta_1)$ can be exactly absorbed into the NLO expansion of NRQCD LDMEs, and in our NLO calculations, we are justified to switch the order of $q_1$-integration from the derivatives and the limit of $q_r$. To achieve this conclusion, we will assume in the following that all distributions of $\zeta_1$ will be convoluted with a function $f(\zeta_1)$, which has a Taylar expansion for the region $-1 < \zeta_1 < 1$.  Since applying the derivative and the limit operations for $q_r$ is equivalent to performing Taylar expansion of $q_r$, in the following, we just compare methods either expanding $q_r$ before or after the $q_1$-integration.

To be specified, we are working at the NLO in the Feynman gauge. Diagrams (a), (b) and (f)-(i) in Fig.~\ref{fig:NLOvirtual} do not cause any problem because they are not connected diagrams and the additional energy-momentum conservation $\delta^D(q_r-q_1)$ makes the integration over $q_1$ trivial. It will be clear later that there is no problem for the diagrams (c), (d), (j), (k) and (l), because these two diagrams do not have Coulomb divergence. After all, we need only to consider carefully the diagram (e), whose amplitude can be written as
\begin{linenomath*}
\begin{align}\label{eq:method2}
\bar{A}(q_1,q_r)=\frac{B(q_1,q_r)}{ \left[\left(q_r-q_1\right)^2+i \varepsilon\right]\left[\left(p/2+q_1\right)^2-m_Q^2+i \varepsilon\right] \left[\left(p/2-q_1\right)^2-m_Q^2+i \varepsilon\right]},
\end{align}
\end{linenomath*}
where $B(q_1,q_r)$ is a polynomial of $q_1$ and $q_r$. In the rest frame of the $\cc$-pair, $p \sim (2m_Q, \vec{0})$, and $q_r \sim (0, m_Q \vec{v})$. In the Coulomb region where $q_1 \sim (m_Q v^2, m_Q \vec{v})$, the relevant scaling relations are: $\left(q_r-q_1\right)^2\sim\left(p/2+q_1\right)^2-m_Q^2 \sim \left(p/2-q_1\right)^2-m_Q^2\sim m_Q^2 v^2$ and $d^Dq_1 \sim m_Q^D v^{D+1}$. Therefore, the leading contribution in this region behaves as $v^{D-5}$, which leads to a $v^{-1}$ Coulomb singularity in four dimensions as $v\to0$.
This simple analysis indicates that the integration region of $q_1 \lesssim q_r$ is indeed very important for the diagram (e).

First, let us consider a simpler case $B(q_1,q_r)=1$.  That is, we need to deal with the following integration,
\begin{linenomath*}
\begin{align}\label{eq:simple}
\begin{split}
I_1\equiv
\int\frac{{\text d}^D q_1}{(2\pi)^D}
\,\frac{\delta(\zeta_1-\frac{2q_1^+}{p^+})}{ \left[\left(q_r-q_1\right)^2+i \varepsilon\right]\left[\left(p/2+q_1\right)^2-m_Q^2+i \varepsilon\right] \left[\left(p/2-q_1\right)^2-m_Q^2+i \varepsilon\right]}
\, .
\end{split}
\end{align}
\end{linenomath*}
%======================
After using the Feynman parametrization to combine the denominators, we have
%======================
\begin{linenomath*}
\begin{align}\label{eq:simpleadd1}
\begin{split}
I_1=\int_0^1 {\text d}x_1
\int_0^{1-x_1} {\text d}y_1
\int\frac{{\text d}^D q_1}{(2\pi)^D}
\frac{2\,\delta(\zeta_1-\frac{2q_1^+}{p^+})}{\big[(q_1-q'_1)^2-\Delta\big]^3}\, ,
\end{split}
\end{align}
\end{linenomath*}
%======================
where
%======================
\begin{linenomath*}
\begin{align}\label{eq:simpleadd2}
\begin{split}
&q'_1=x_1q_r+(1-x_1-2y_1)\,p/2,
\\
&
\Delta=(1-x_1)^2 q_r^2+(1-x_1-2y_1)^2 \,p^2/4 -i\varepsilon.
\end{split}
\end{align}
\end{linenomath*}
%======================
In Eq.~(\ref{eq:simpleadd1}), with a single pole, the integration of $q_1^-$ vanishes unless $q_1^+={q'}_1^+$. In general, the following relation,
%======================
\begin{linenomath*}
\begin{align}\label{eq:simpleadd3}
\begin{split}
\int\frac{{\text d}^D q_1}{(2\pi)^D}
\frac{\delta(\zeta_1-\frac{2q_1^+}{p^+})}{\big[(q_1-q'_1)^2-\Delta\big]^n}
=
\delta(\zeta_1-\frac{2{q'}_1^+}{p^+})
\int\frac{{\text d}^D q_1}{(2\pi)^D}
\frac{1}{\big[(q_1-q'_1)^2-\Delta\big]^n}\, ,
\end{split}
\end{align}
\end{linenomath*}
%======================
is valid when both sides are convoluted with any smooth function $f(\zeta_1)$ that can be Taylor expanded in the region $-1<\zeta_1<1$.  Applying the relation in Eq.~(\ref{eq:simpleadd3}) to the integration in Eq.~(\ref{eq:simpleadd1}), and performing the $q_1$-integration, we obtain
%======================
\begin{linenomath*}
\begin{align}\label{eq:simpleadd4}
\begin{split}
I_1 =
- \frac{i}{(4\pi)^{2-\epsilon}} \frac{\Gamma(1+\epsilon)}{(p^2/4)^{1+\epsilon}}
\int_0^1 {\text d}x_1
\int_0^{1-x_1} {\text d}y_1
\frac{\delta(\zeta_1-2{q'}_1^+/p^+)}{\Delta^{1+\epsilon}}\, ,
\end{split}
\end{align}
\end{linenomath*}
%======================
where $\epsilon=(4-D)/2$.   By changing variable $x_1=1-x$ and then letting $y_1=x(1-y)/2$,
we can rewrite $I_1$ as
%======================
\begin{linenomath*}
\begin{align}\label{eq:simple1}
\begin{split}
I_1 =
- \frac{i}{(4\pi)^{2-\epsilon}}\,
\frac{\Gamma(1+\epsilon)}{(p^2/4)^{1+\epsilon}}\,
\frac{1}{2}\, Z(\zeta_1,q_r),
\end{split}
\end{align}
\end{linenomath*}
with
\begin{linenomath*}
\begin{align}
\begin{split}\label{eq:Zqr}
Z(\zeta_1,q_r)=\int_{-1}^1 \frac{dy}{\left(y^2 - \beta^2-i \varepsilon\right)^{1+\epsilon}}
\int_0^1 \frac{dx}{x^{1+2\epsilon}} \delta(\zeta_1 - \tilde{\beta} + x y + x\tilde{\beta}),
\end{split}
\end{align}
\end{linenomath*}
%======================
where $\tilde{\beta}=2q_r^+/p^+$ and $\beta^2=-4q_r^2/p^2$, and both are small parameters.
Since $Z(\zeta_1,q_r)$ will eventually convolute with a non-singular function $f(\zeta_1)$,
we can expand the $\delta$-function as
\begin{linenomath*}
\begin{align}\label{eq:delta-zeta}
\begin{split}
&\delta(\zeta_1 - \tilde{\beta} + x y + x\tilde{\beta})\\
=& \delta(\zeta_1) + \delta'(\zeta_1)(- \tilde{\beta} + x y + x\tilde{\beta}) + \cdots\\
=& \sum_{i,j\ge k,k} C_{i,j,k} \tilde{\beta}^i x^j y^k\\
=& \sum_{i,j\ge 2k,k} C_{i,j,2k} \tilde{\beta}^i x^j y^{2k},
\end{split}
\end{align}
\end{linenomath*}
where $i$, $j$ and $k$ are natural numbers, and the power of $x$ cannot be less than the power of $y$.  The equation in Eq.~(\ref{eq:delta-zeta}) is a result of the fact that terms odd in $y$ vanish under the integration of $y$ from $-1$ to $1$. Then, the $x$-integration in $Z(\zeta_1,q_r)$ is trivial,
\begin{linenomath*}
\begin{align}
\begin{split}
\int_0^1 \frac{dx}{x^{1+2\epsilon}} x^j = \frac{1}{j-2\epsilon}.
\end{split}
\end{align}
\end{linenomath*}
To perform the $y$-integration, we introduce a parameter $\Lambda \gg \beta$, and rewrite the $y$-integration as
\begin{linenomath*}
\begin{align}\label{eq:y-integral}
\begin{split}
\int_{-1}^1 \frac{y^{2k}dy}{\left(y^2-\beta^2-i \varepsilon\right)^{1+\epsilon}}
=\left(\int_{-1}^{-\Lambda}+\int_\Lambda^1\right)\frac{y^{2k}dy}
{\left(y^2-\beta^2-i \varepsilon\right)^{1+\epsilon}}
+ \int_{-\Lambda}^\Lambda\frac{y^{2k}dy}{\left(y^2-\beta^2-i \varepsilon\right)^{1+\epsilon}}\, .
\end{split}
\end{align}
\end{linenomath*}
Since $y^2 \ge \Lambda^2 \gg \beta^2$ in the first term above, we can expand $\beta^2$
before performing the $y$-integration, and obtain
\begin{linenomath*}
\begin{align}
\begin{split}
 \frac{y^{2k}}{\left(y^2-\beta^2-i \varepsilon\right)^{1+\epsilon}}
 =\frac{y^{2k}}{y^{2-2\epsilon}}+(1+\epsilon)\frac{y^{2k}}{y^{4+2\epsilon}} \beta^2
 + \cdots \equiv E_k(y^2)\, ,
\end{split}
\end{align}
\end{linenomath*}
and
\begin{linenomath*}
\begin{align}
\begin{split}
\int_{-1}^1 \frac{y^{2k}dy}{\left(y^2-\beta^2-i \varepsilon\right)^{1+\epsilon}}
=&\left(\int_{-1}^{-\Lambda}+\int_\Lambda^1\right)E_k(y^2) dy
+ \int_{-\Lambda}^\Lambda\frac{y^{2k}dy}{\left(y^2-\beta^2-i \varepsilon\right)^{1+\epsilon}}\, .
\end{split}
\end{align}
\end{linenomath*}
This identity can also be written as
\begin{linenomath*}
\begin{align}
\begin{split}
\int_{-1}^1 \frac{y^{2k}dy}{\left(y^2-\beta^2-i \varepsilon\right)^{1+\epsilon}}
- \int_{-1}^1 E_k(y^2) dy
=
\int_{-\Lambda}^\Lambda\frac{y^{2k}dy}{\left(y^2-\beta^2-i \varepsilon\right)^{1+\epsilon}}
- \int _{-\Lambda}^\Lambda E_k(y^2) dy \, .
\end{split}
\end{align}
\end{linenomath*}
In the LHS of above identity, the first term corresponds to performing $y$-integration before expanding $\beta^2$, while the second term corresponds to expanding $\beta^2$ before doing the $y$-integration.  The RHS provides the corrections to the original $y$-integration caused by expanding $\beta^2$ first.  Since the corrections on the RHS does not depend on the choice of $\Lambda$ as long as $\Lambda \gg \beta$, we can choose $\Lambda=\infty$ to simplify the identity as,
\begin{linenomath*}
\begin{align}\label{eq:identity}
\begin{split}
\int_{-1}^1 \frac{y^{2k}dy}{\left(y^2-\beta^2-i \varepsilon\right)^{1+\epsilon}}
-\int_{-1}^1 E_k(y^2) dy
=&\int_{-\infty}^{+\infty}\frac{y^{2k}dy}{\left(y^2-\beta^2-i \varepsilon\right)^{1+\epsilon}}\\
=&\beta^{2k-1-2\epsilon}\int_{-\infty}^{+\infty}\frac{y^{2k}dy}{\left(y^2-1-i \varepsilon\right)^{1+\epsilon}}.
\end{split}
\end{align}
\end{linenomath*}
In deriving above simplified identify, we used
\begin{linenomath*}
\begin{align}
\begin{split}
\int_{-\infty}^{+\infty} E_k(y^2) dy = \int_{-\infty}^{+\infty}
\left[ \frac{y^{2k}}{y^{2+2\epsilon}}+(1+\epsilon)\frac{y^{2k}}{y^{4+2\epsilon}} \beta^2 + \cdots \right] dy =0.
\end{split}
\end{align}
\end{linenomath*}
Note that using dimensional regularization is crucial for deriving above results.   Although the integration on the RHS of Eq.~(\ref{eq:identity}) could be further carried out, its result is not really relevant for our discussion here. Instead, we need to point out that it is an odd function of $\beta$.

In comparison with the situation discussed in Ref.~\cite{Beneke:1997zp}, the second term on the LHS of Eq.~(\ref{eq:identity}) corresponds to contributions from hard region, while the term on the RHS of the equation corresponds to contributions from potential region, which can be exactly reproduced by NLO calculation of NRQCD LDMEs. Note also that deriving Eq.~\eqref{eq:simple1} from Eq.~\eqref{eq:simple} by performing Feynman parametrization and integrating out $q_1$ did not miss anything. Therefore, we conclude that, if we are not interested in the contributions from potential region, we can calculate Eq.~\eqref{eq:simple} by expanding the $q_r$ before doing the $q_1$-integration.

When $B(q_1,q_r)$ is a general polynomial of $q_1$ and $q_r$, we can carry out essentially all steps in our arguments above for the situation when $B(q_1,q_r)=1$.  We can still expand the $\delta$-function, use the Feynman parametrization to re-organize the $q_1$-integral, and perform the integration of $q_1$ before integration over Feynman parameters. The key difference is that we get a slightly different $y$-integral,
\begin{linenomath*}
\begin{align}
\begin{split}
\int_{-1}^1 \frac{y^{2k}dy}{\left(y^2+\beta^2-i \varepsilon\right)^{d+1+\epsilon}} ,
\end{split}
\end{align}
\end{linenomath*}
where $d$ is an integer. The trick of introducing a $\Lambda\gg \beta$ is still valid for showing that expanding the $q_r$ before doing the $q_1$-integration is effectively neglecting the Coulomb region.  Since the Coulomb region is cancelled exactly by NLO calculation of NRQCD LDMEs, we conclude that we can get correct short-distance coefficients at NLO if we expand $q_r$ before the integration of $q_1$.

%%%%%%%%%%%%%%%%%%%%%%%%%%%%%%%%%
%                                                   						  %
%                                   App II: app_results				  %
%  											   	  %
%%%%%%%%%%%%%%%%%%%%%%%%%%%%%%%%%

\section{Results of fragmentation functions to $P$-wave heavy quark pair}\label{app:results}

In this Appendix, we summarize our results of short-distance coefficients for NRQCD factorization
expansion of heavy quark-pair FFs to a heavy quarkonium through all possible $P$-wave states of
a non-relativistic heavy quark pair.

\subsection{Definitions and Notations}

As a conjecture, we factorize the heavy quarkonium FFs from a heavy quark pair
at the input factorization scale, $\mu_0\gtrsim 2m_Q$, in a NRQCD factorization formalism,
\begin{align}
\begin{split}\label{eq:NRQCDFacDouble}
{\mathcal D}_{[\cc(\kappa)]\to H}(z,\zeta_1,\zeta_2;m_Q,\mu_0)
&=
\sum_{[\cc(n)]}
\Big\{\hat{d}^{\,(0)}_{[\cc(\kappa)]\to [\cc(n)]}(z,\zeta_1,\zeta_2;m_Q,\mu_0,\mu_\Lambda)
\\
&\hspace{-3cm}+
\left(\frac{\as}{\pi}\right)
\hat{d}^{\,(1)}_{[\cc(\kappa)]\to [\cc(n)]}(z,\zeta_1,\zeta_2;m_Q,\mu_0,\mu_\Lambda)
+O(\alpha_s^2)\Big\}
\times
\frac{\langle \mathcal{O}_{[\cc(n)]}^{H}(\mu_\Lambda)\rangle}{m_Q^{2L+1}},
\end{split}
\end{align}
where $[\cc(\kappa)]$ labels a fragmenting heavy quark pair, perturbatively produced in high energy hard collisions, and the $\kappa$ represents the pair's spin-color states, which can be a vector ($v$), axial-vector ($a$) or tensor ($t$) spin state, with either color singlet or octet.
$H$ is a physical heavy quarkonium, such as $h_Q$ and $\chi_{QJ}$, with $Q$ being charm or bottom quark.
$[\cc(n)]$ labels all intermediate NRQCD states with $n$ usually expressed as $^{2S+1}L_J^{[1,8]}$.
Color singlet LDMEs could be related to heavy quarkonia wave functions at the origin up to a relative normalization, such as
\begin{align}
%==============1P1==============
\begin{split}\label{eq:LDME1P11}
\langle \mathcal{O}_{[\cc(\CSaPa)]}^{\ h_Q}\rangle
=
\frac{9}{4\pi}|R'_{ h_Q}(0)|^2,
\end{split}
\\
%==============3PJ==============
\begin{split}\label{eq:LDME3PJ1}
\langle \mathcal{O}_{[\cc(\CScPj)]}^{\ \chi_{QJ}}\rangle
=
\frac{3(2J+1)}{4\pi}|R'_{ \chi_{QJ}}(0)|^2.
\end{split}
\end{align}
On the other hand, the color octet LDMEs could only be extracted from data.
In the rest of this Appendix, we list both the LO and NLO short-distance coefficients,
$\hat{d}^{\,(0)}$ and $\hat{d}^{\,(1)}$, in Eq.~(\ref{eq:NRQCDFacDouble}).

%%%%%%%%%%%%%%%%%%%%%%%%%%%
%%     subsection: LO
%%%%%%%%%%%%%%%%%%%%%%%%%%%
\subsection{LO results} \label{sec:LOresults}

In this part we list all non-vanishing LO short-distance coefficents.
\begin{align}
%==========V1  to  3p0[1]===========
\begin{split}\label{eq:v1to3p01LO}
\hat{d}^{\text{(0)}}_{\vone \to \CScPz}(\zeta_1,\zeta_2,z)
=\frac{1}{2(3-2\epsilon)}\delta'(\zeta_1)\delta'(\zeta_2)\delta(1-z),
\end{split}\\
%==========V1  to  3p2[1]===========
\begin{split}\label{eq:v1to3p21LO}
\hat{d}^{\text{(0)}}_{\vone \to \CScPb}(\zeta_1,\zeta_2,z)
=\frac{1}{(3-2\epsilon)(5-2\epsilon)}\delta'(\zeta_1)\delta'(\zeta_2)\delta(1-z),
\end{split}\\
%==========A1  to  1p1[1]===========
\begin{split}\label{eq:a1to1p11LO}
\hat{d}^{\text{(0)}}_{\aone \to \CSaPa}(\zeta_1,\zeta_2,z)
=\frac{1}{2(3-2\epsilon)}\delta'(\zeta_1)\delta'(\zeta_2)\delta(1-z),
\end{split}\\
%==========A1  to  3p1[1]===========
\begin{split}\label{eq:a1to3p11LO}
\hat{d}^{\text{(0)}}_{\aone \to \CScPa}(\zeta_1,\zeta_2,z)
=\frac{(1-2\epsilon)}{(3-2\epsilon)}\delta(\zeta_1)\delta(\zeta_2)\delta(1-z),
\end{split}\\
%==========T1  to  1p1[1]===========
\begin{split}\label{eq:t1to1p11LO}
\hat{d}^{\text{(0)}}_{\tone \to \CSaPa }(\zeta_1,\zeta_2,z)
=\frac{(1-2\epsilon)}{2(3-2\epsilon)}\delta(\zeta_1)\delta(\zeta_2)\delta(1-z),
\end{split}\\
%==========T1  to  3p1[1]===========
\begin{split}\label{eq:t1to3p11LO}
\hat{d}^{\text{(0)}}_{\tone \to \CScPa}(\zeta_1,\zeta_2,z)
=\frac{1}{2(3-2\epsilon)(2-2\epsilon)}\delta'(\zeta_1)\delta'(\zeta_2)\delta(1-z),
\end{split}\\
%==========T1  to  3p2[1]===========
\begin{split}\label{eq:t1to3p21LO}
\hat{d}^{\text{(0)}}_{\tone \to \CScPb}(\zeta_1,\zeta_2,z)
=\frac{1}{2(5-2\epsilon)(2-2\epsilon)}\delta'(\zeta_1)\delta'(\zeta_2)\delta(1-z),
\end{split}\\
%==========s1 8  to  s2 8===========
\begin{split}\label{eq:s18tos28LO}
\hat{d}^{\text{(0)}}_{s^{[8]}\to \state{{2S+1}}{P}{J}{8}}(\zeta_1,\zeta_2,z)
=\frac{1}{N_c^2-1}\hat{d}^{\text{(0)}}_{s^{[1]}\to \state{{2S+1}}{P}{J}{1}}(\zeta_1,\zeta_2,z),
\end{split}
%========The End==============
\end{align}
where $s$ in the last equation could be $v$, $a$ or $t$, and the dimension is defined as $D=4-2\epsilon$.

%%%%%%%%%%%%%%%%%%%%%%%%%%%%%%%%%%
%                                                                                                       %
%                       Delta Functions                                                       %
%                                                                                                      %
%%%%%%%%%%%%%%%%%%%%%%%%%%%%%%%%%%
\subsection{$\Delta$-functions}

To better present the NLO results, we need some auxiliary functions, i.e. $\Delta$-functions in this subsection and generalized ``$\pm$'' distributions in next subsection, to summarize the general structure of $\zeta_1$ and $\zeta_2$ dependence of the short-distance coefficients.
As discussed in section \ref{sec:general}, the possible structures of $\delta$-functions are very limited.
In this subsection, we list all of them and their asymptotic behaviors as $z\to1$.
%
%===========================
\begin{align}
%==========
\begin{split}
\DeltaZero
=
4\,\delta(\zeta_1)\delta(\zeta_2),
\end{split}\\
% 0 PP
\begin{split}
\DeltaPPZero
=
4 \,z^2\,\delta'(\zeta_1)\delta'(\zeta_2),
\end{split}\\
% 1
\begin{split}
\DPMOne
=4\left[\DAA \pm \DBB\right] \left[\DXX \pm \DYY\right],
\end{split}\\
% 1 P
\begin{split}
&\DPMPOne
=-4\,z\Big\{
\left[\DPAA \pm \DPBB\right] \left[\DXX \pm \DYY\right]
\\
&\hspace{1cm}
+\left[\DAA\pm\DBB\right] \left[\DPXX\pm\DPYY\right]
\Big\},
\end{split}\\
% 1 PP
\begin{split}
\DPMPPOne
=4\,z^2\,\left[\DPAA \pm \DPBB\right] \left[\DPXX \pm \DPYY\right],
\end{split}\\
% 8
\begin{split}
&\DPMEight
=4\Big\{(N_c^2-2)\left[\DAA\DXX+\DBB\DYY\right]\\
&\hspace{1cm}
\mp 2\left[\DAA\DYY+\DBB\DXX\right] \Big\},
\end{split}\\
% 8 P
\begin{split}
&\DPMPEight
= -4\, z\,\Big\{(N_c^2-2)\big[\DPAA\DXX+\DAA\DPXX
\\
&\hspace{1cm}
+\DPBB\DYY+\DBB\DPYY\big]
\\
&\hspace{1cm}
\mp 2\big[\DPAA\DYY+\DAA\DPYY
\\
&\hspace{1cm}
+\DPBB\DXX+\DBB\DPXX\big]
\Big\},
\end{split}\\
% 8 PP
\begin{split}
&\DPMPPEight
=4\,z^2\,\Big\{(N_c^2-2)\left[\DPAA\DPXX+\DPBB\DPYY\right]\\
&\hspace{1cm}
\mp 2\left[\DPAA\DPYY+\DPBB\DPXX\right] \Big\},
\end{split}
\end{align}
All these $\Delta$-functions are invariant under the transformation ($\zeta_1 \to -\zeta_1$, $\zeta_2 \to -\zeta_2$) and the exchange $\zeta_1 \leftrightarrow \zeta_2$, including the crossing exchange
($\zeta_1 \to -\zeta_2$, $\zeta_2 \to -\zeta_1$).
In addition, $\DeltaZero$, $\DPlusOne$, $\DPlusPOne$ and $\DPlusPPOne$ are even in both $\zeta_1$ and $\zeta_2$, while $\DeltaPPZero$, $\DMinusOne$, $\DMinusPOne$ and $\DMinusPPOne$ are odd in both $\zeta_1$ and $\zeta_2$.
Under the integration of $\zeta_1$ and $\zeta_2$ with a well behaved test function, the asymptotic behaviors of these $\Delta$-functions at $z\to1$ are
%
%==========================
\begin{align}
%---------------- begin -----------------
\begin{split}
\lim_{z \to 1}\DPlusOne =O[1],
\hspace{3.15cm}
\lim_{z \to 1}\DMinusOne =O[(1-z)^2],
\end{split}
\nonumber
\\
%%%
\begin{split}
\lim_{z \to 1}\DPlusPOne =O[(1-z)],
\hspace{2cm}
\lim_{z \to 1}\DMinusPOne =O[(1-z)],
\end{split}
\nonumber
\\
%%%
\begin{split}
\lim_{z \to 1}\DPlusPPOne =O[(1-z)^2],
\hspace{1.75cm}
\lim_{z \to 1}\DMinusPPOne =O[1],
\end{split}
\nonumber
\\
%%%
\begin{split}
\lim_{z \to 1}\DPMEight =O[1],
\hspace{3.05cm}
\lim_{z \to 1}\DPMPEight =O[(1-z)],
\end{split}
\nonumber
\\
%%%
\begin{split}
\lim_{z \to 1}\DPMPPEight =O[1],
\end{split}
\nonumber
\end{align}
%==========================
Therefore,
%==========================
\begin{align}
\begin{split}
\frac{\DMinusOne}{(1-z)},
\hspace{0.3cm}
\frac{\DPMPOne}{(1-z)},
\hspace{0.3cm}
\frac{\DPlusPPOne}{(1-z)},
\hspace{0.3cm}
\text{and}
\hspace{0.2cm}
\frac{\DPMPEight}{(1-z)}\nonumber
\end{split}
\end{align}
%==========================
do not exhibit any pole at $z=1$.

%%%%%%%%%%%%%%%%%%%%%%%%%%%%%%%%%%
%                                                                                                       %
%                       plus minus distribution                                            %
%                                                                                                      %
%%%%%%%%%%%%%%%%%%%%%%%%%%%%%%%%%%

\subsection{Generalized Plus-distributions and Minus-distributions}

We define generalized plus and minus distributions to regularize the singularities at $\zeta_1=0$ and $\zeta_2=0$.
They are  collectively defined as
%==========================
\begin{linenomath*}
\begin{align}\label{eq:plusminusalldef}
\begin{split}
\Big(g(\zeta_1)\Big)_{m\pm}
\equiv
\int_{-1}^{1}\left[\theta(x) \pm \theta(-x)\right] g(|x|)
\times
\left(\delta(x-\zeta_1)-\sum_{i=0}^{m-1}\frac{\delta^{(i)}(\zeta_1)}{i\,!}(-x)^{\,i}\right)dx\, ,
\end{split}
\end{align}
\end{linenomath*}
%==========================
where $\delta^{(i)}(\zeta_1)$ represents the $i$-th derivative of the $\delta$-function.
More explicitly, plus and minus distributions have the following relation under the integration with a test function,
%==========================
\begin{linenomath*}
\begin{align}\label{eq:plusminusall}
\begin{split}
\int \Big(g(\zeta_1)\Big)_{m\pm}f(\zeta_1)\,d\zeta_1
\equiv
\int_{-1}^{1}\left[\theta(\zeta_1) \pm \theta(-\zeta_1)\right] g(|\zeta_1|)
\times
\left(f(\zeta_1)-\sum_{i=0}^{m-1}\frac{f^{(i)}(0)}{i\,!}\zeta_1^{\,i}\right)d\zeta_1.
\end{split}
\end{align}
\end{linenomath*}
%==========================
From the above definition, we find,
\begin{linenomath*}
\begin{align}\label{eq:plusminusparity}
\begin{split}
\Big(g(-\zeta_1)\Big)_{m\pm}
&=
\int_{-1}^{1}\left[\theta(x) \pm \theta(-x)\right] g(|x|)
\times
\left(\delta(x+\zeta_1)-\sum_{i=0}^{m-1}\frac{\delta^{(i)}(-\zeta_1)}{i\,!}(-x)^{\,i}\right)dx\\
&=
\int_{-1}^{1}\left[\theta(-x) \pm \theta(x)\right] g(|x|)
\times
\left(\delta(-x+\zeta_1)-\sum_{i=0}^{m-1}\frac{\delta^{(i)}(-\zeta_1)}{i\,!}(x)^{\,i}\right)dx\\
&=
\int_{-1}^{1} \pm \left[\theta(x) \pm \theta(-x)\right] g(|x|)
\times
\left(\delta(x-\zeta_1)-\sum_{i=0}^{m-1}\frac{\delta^{(i)}(\zeta_1) }{i\,!}(-x)^{\,i}\right)dx\\
&=\pm \Big(g(\zeta_1)\Big)_{m\pm}.
\end{split}
\end{align}
\end{linenomath*}
That is, plus function is an even function with respect to $\zeta_1$, and minus is an odd function with respect to $\zeta_1$. For $m\ge0$, we have
%==========================
\begin{linenomath*}
\begin{align}
\begin{split}
\Big(g(\zeta_1)\Big)_{(2m+2)+}
=&
\int_{-1}^{1}\left[\theta(x) + \theta(-x)\right] g(|x|)
\times
\left(\delta(x-\zeta_1)-\sum_{i=0}^{2m+1}\frac{\delta^{(i)}(\zeta_1)}{i\,!}(-x)^{\,i}\right)dx\\
=& \Big(g(\zeta_1)\Big)_{(2m+1)+} -
\int_{-1}^{1}\left[\theta(x) + \theta(-x)\right] g(|x|)
\times
\frac{\delta^{(2m+1)}(\zeta_1)}{(2m+1)\,!}(-x)^{\,2m+1} dx\\
=& \Big(g(\zeta_1)\Big)_{(2m+1)+},
\end{split}
\end{align}
\end{linenomath*}
%==========================
and
%==========================
\begin{linenomath*}
\begin{align}
\begin{split}
\Big(g(\zeta_1)\Big)_{(2m+1)-}
=&
\int_{-1}^{1}\left[\theta(x) - \theta(-x)\right] g(|x|)
\times
\left(\delta(x-\zeta_1)-\sum_{i=0}^{2m}\frac{\delta^{(i)}(\zeta_1)}{i\,!}(-x)^{\,i}\right)dx\\
=& \Big(g(\zeta_1)\Big)_{(2m)-} -
\int_{-1}^{1}\left[\theta(x) - \theta(-x)\right] g(|x|)
\times
\frac{\delta^{(2m)}(\zeta_1)}{(2m)\,!}(-x)^{\,2m} dx\\
=& \Big(g(\zeta_1)\Big)_{(2m)-}.
\end{split}
\end{align}
\end{linenomath*}
%==========================

%%%%%%%%%%%%%%%%%%%%%%%%%%%%%%%%%%
%                                                                                                       %
%                       NLO results for v                                                     %
%                                                                                                      %
%%%%%%%%%%%%%%%%%%%%%%%%%%%%%%%%%%

\subsection{P-wave NLO results with an initial vector $\cc$-state}

In this subsection, we list results of NLO short-distance coefficients to the fragmentation functions for a vector pQCD $\cc$-state to fragment into a $P$-wave NRQCD $\cc$-state.
Fragmentation channels that are equal to zero at this order are not listed.

\begin{align}
%==========V1  to  3P0[1]===========
\begin{split}\label{eq:v1to3p01NLO}
&\hat{d}^{\text{(1)}}_{\vone\to \CScPz}
=-\frac{1}{12 }C_F\delta(1-z)
\Big\{
\frac{\DeltaPPZero}{4}(\LogUV-4)
\\
&\hspace{2cm}
+{\tilde{V}'}_{va}(\zeta_1,\zeta_2)(\LogUV-\frac{2}{3})
+{V_{v1}'}(\zeta_1,\zeta_2)
\Big\}
,
\end{split}\\
%==========V1  to  3P2[1]===========
\begin{split}\label{eq:v1to3p21NLO}
&\hat{d}^{\text{(1)}}_{\vone\to \CScPb}
=-\frac{1}{30  }C_F\delta(1-z)
\Big\{
\frac{\DeltaPPZero}{4}(\LogUV-4)
\\
&\hspace{2cm}
+{\tilde{V}'}_{va}(\zeta_1,\zeta_2)(\LogUV-\frac{16}{15})
+{V_{v2}'}(\zeta_1,\zeta_2)
\Big\}
,
\end{split}\\
%==========V1  to  1P1[8]===========
\begin{split}\label{eq:v1to1p18NLO}
&\hat{d}^{\text{(1)}}_{\vone\to \COaPa}
=\frac{1}{12}\frac{C_F}{(N_c^2-1)}z(1-z)
\Big\{
\big[\DMinusPPOne+\DMinusPOne +\DMinusOne\big] (\frac{1}{2}\LogUV-\frac{1}{3})
\\
&\hspace{2cm}
-\DMinusPPOne\big(\mylog{2-2z}+\frac{7}{6}\big) -\DMinusPOne\big(\mylog{2-2z}+\frac{5}{3}\big)
\\
&\hspace{2cm}
-\DMinusOne\big(\mylog{2-2z}+\frac{7}{6}\big)
\Big\}
,
\end{split}\\
%==========V1  to  3P0[8]===========
\begin{split}\label{eq:v1to3p08NLO}
&\hat{d}^{\text{(1)}}_{\vone\to \COcPz}
=\frac{z}{12}\frac{C_F}{(N_c^2-1)}
\Big\{
\frac{2}{3}\DeltaZero\,\delta(1-z)(-\LogIR+2\logtwo)
\\
&\hspace{2cm}
+\big[\frac{\DPlusPPOne}{(1-z)} +\DPlusPOne +\DPlusOne(1-z)\big] (\frac{1}{2}\LogUV-\frac{1}{3})
\\
&\hspace{2cm}
-\frac{\DPlusPPOne}{(1-z)}  R_{v1}(z)- \frac{\DPlusPOne}{(1-z)} R_{v2}(z)-\DPlusOne \,R_{v3}(z)
\Big\}
,
\end{split}\\
%==========V1  to  3P1[8]===========
\begin{split}\label{eq:v1to3p18NLO}
&\hat{d}^{\text{(1)}}_{\vone\to \COcPa}
=\frac{z}{12}\frac{C_F}{(N_c^2-1)}
\Big\{
\frac{2}{3}\DeltaZero\,\delta(1-z)(-\LogIR+2\logtwo+\frac{1}{2})
\\
&\hspace{2cm}
+\DPlusOne (1-z) (\LogUV+\frac{4}{3})
+\frac{\DPlusPPOne}{2}(1-z)+ \frac{\DPlusPOne}{2}(\frac{3}{2}-z)
\\
&\hspace{2cm}
+\DPlusOne\,\big[\frac{1}{3}\frac{1}{(1-z)_+}-2(1-z)\mylog{2-2z}+\frac{3}{2}z-\frac{7}{6}\big]
\Big\}
,
\end{split}\\
%==========V1  to  3P2[8]===========
\begin{split}\label{eq:v1to3p28NLO}
&\hat{d}^{\text{(1)}}_{\vone\to \COcPb}
=\frac{z}{30 }\frac{C_F}{(N_c^2-1)}
\Big\{
\frac{5}{3}\DeltaZero\,\delta(1-z)(-\LogIR+2\logtwo+\frac{3}{10})
\\
&\hspace{2cm}
+\big[\frac{\DPlusPPOne}{(1-z)}+\DPlusPOne+\DPlusOne(1-z)\big] (\frac{1}{2}\LogUV-\frac{8}{15})
\\
&\hspace{2cm}
-\frac{\DPlusPPOne}{(1-z)}  R_{v4}(z)- \frac{\DPlusPOne}{(1-z)} R_{v5}(z)-\DPlusOne \,R_{v6}(z)
\Big\}
,
\end{split}\\
%==========V8  to  1P1[8]===========
\begin{split}\label{eq:v8to1p18NLO}
&\hat{d}^{\text{(1)}}_{\veight\to \COaPa}
=\frac{1}{12}\frac{C_F}{(N_c^2-1)^2}z(1-z)
\Big\{
\big[\DPlusPPEight +\DPlusPEight +\DPlusEight\big] (\frac{1}{2}\LogUV-\frac{1}{3})
\\
&\hspace{2cm}
-\DPlusPPEight\big(\mylog{2-2z}+\frac{7}{6}\big)
-\DPlusPEight\big(\mylog{2-2z}+\frac{5}{3}\big)
\\
&\hspace{2cm}
-\DPlusEight\big(\mylog{2-2z}+\frac{7}{6}\big)
\Big\}
,
\end{split}\\
%==========V8  to  3P0[8]===========
\begin{split}\label{eq:v8to3p08NLO}
&
\hat{d}^{\text{(1)}}_{\veight \to \COcPz}
=
\frac{z}{12}
\frac{C_F}{(N_c^2-1)^2}
\Big\{
\delta(1-z)
\Big[\,
\frac{1}{3}(N_c^2-4)\DeltaZero\Big(-\LogIR+2\logtwo\Big)
\\
&\hspace{2cm}
+\frac{1}{4}\DeltaPPZero\Big(c \times \LogUV+c_{1}\Big)
+\tilde{V}_{va}'(\zeta_1,\zeta_2)\big(\LogUV-\frac{2}{3}\,\big)
+V'_{v1}(\zeta_1,\zeta_2)
\Big]
\\
&\hspace{2cm}
+\Big[\frac{\DMinusPPEight}{(1-z)_+}+{\DMinusPEight} +{\DMinusEight} (1-z)
\Big]\big(\,\frac{1}{2}\LogUV-\frac{1}{3}\,\big)
\\
&\hspace{2cm}
-
\DMinusPPEight\,R_{v7}(z)-\frac{\DMinusPEight}{(1-z)}R_{v2}(z)-\DMinusEight R_{v3}(z)
\Big\}
,
\end{split}\\
%==========V8  to  3P1[8]===========
\begin{split}\label{eq:v8to3p18NLO}
&\hat{d}^{\text{(1)}}_{\veight\to \COcPa}
=\frac{z}{12  }\frac{C_F}{(N_c^2-1)^2}
\Big\{
\frac{1}{3}(N_c^2-4)\DeltaZero\,\delta(1-z)(-\LogIR+2\logtwo+\frac{1}{2})
\\
&\hspace{2cm}
+\DMinusEight (1-z) (\LogUV+\frac{4}{3})
+\frac{\DMinusPPEight}{2}(1-z)+\frac{\DMinusPEight}{2}(\frac{3}{2}-z)
\\
&\hspace{2cm}
+\DMinusEight\,\big[\frac{1}{3}\frac{1}{(1-z)_+}-2(1-z)\mylog{2-2z}+\frac{3}{2}z-\frac{7}{6}\big]
\Big\}
,
\end{split}\\
%==========V8  to  3P2[8]===========
\begin{split}\label{eq:v8to3p28NLO}
&
\hat{d}^{\text{(1)}}_{\veight \to \COcPb}
=
\frac{z}{30}
\frac{C_F}{(N_c^2-1)^2}
\Big\{
\delta(1-z)
\Big[\,
 \frac{5}{6}(N_c^2-4)\DeltaZero\Big(-\LogIR+2\logtwo+\frac{3}{10}\Big)
\\
&\hspace{2cm}
+\frac{1}{4}\DeltaPPZero\Big(c \times \LogUV+c_{1}\Big)
+\tilde{V}_{va}'(\zeta_1,\zeta_2)\big(\LogUV-\frac{16}{15}\,\big)
+V'_{v2}(\zeta_1,\zeta_2)
\Big]
\\
&\hspace{2cm}
+
\Big[
\frac{\DMinusPPEight}{(1-z)_+}+{\DMinusPEight}+{\DMinusEight} (1-z)
\Big]\big(\, \frac{1}{2}\LogUV-\frac{8}{15}\,\big)
\\
&\hspace{2cm}
-
\DMinusPPEight\,R_{v8}(z)-\frac{\DMinusPEight}{(1-z)}R_{v5}(z)-\DMinusEight R_{v6}(z)
\Big\}
,
\end{split}\\
%==========s8  to  s[1]===========
\begin{split}\label{eq:s8tos1NLO}
\hat{d}^{\text{(1)}}_{v^{[8]} \to \state{{2S+1}}{L}{J}{1}}
=\hat{d}^{\text{(1)}}_{v^{[1]} \to \state{{2S+1}}{L}{J}{8}},
\end{split}
%========The End==============
\end{align}
where the dependence on $z$, $\zeta_1$, $\zeta_2$, and $\mu_F$ in the last equation is suppressed.
$\tilde{V}$, $V$, $R$ and $c$ functions in above equations are defined as
%
%======== Aux functions =============
\begin{align}
%=========\V'_{va}==========
\begin{split}
{\tilde{V}'}_{va}(\zeta_1,\zeta_2)
=
\delta'(\zeta_2)\Big\{
\Minusb-\frac{1}{2}\Minusz{\zeta_1+1}
\Big\}
+(\zeta_1 \leftrightarrow \zeta_2)
,
\end{split}\\
%=========V'_{1}==========
\begin{split}
&{V_{v1}}'(\zeta_1,\zeta_2)
=
\delta'(\zeta_2)
\Big\{
\Minusc
-\MinusLogb+\frac{5}{3}\Minusb+\Minusa
\\
&\hspace{2cm}
+\frac{1}{2}\Minusz{(\zeta_1+1)\mylog{\zeta_1^2}}
+\frac{1}{6}\Minusz{\zeta_1-23}
\Big\}
+(\zeta_1 \leftrightarrow \zeta_2)
,
\end{split}\\
%=========V'_{2}==========
\begin{split}
&{V_{v2}}'(\zeta_1,\zeta_2)
=
\delta'(\zeta_2)
\Big\{
\Minusc
-\MinusLogb+\frac{31}{15}\Minusb-\frac{7}{2}\Minusa
\\
&\hspace{2cm}
+\frac{1}{2}\Minusz{(\zeta_1+1)\mylog{\zeta_1^2}}
+\frac{22}{15}\Minusz{\zeta_1-\frac{31}{44}}
\Big\}
+(\zeta_1 \leftrightarrow \zeta_2)
,
\end{split}\\
%=========R v1==========
\begin{split}
&
R_{v1}(z)
=
\mylog{2-2z}+\frac{1}{6}
\,,
\end{split}\\
%=========R v2==========
\begin{split}
&
R_{v2}(z)
=
(1-z)\mylog{2-2z}+\frac{1}{3}z+\frac{1}{6}
\,,
\end{split}\\
%=========R v3==========
\begin{split}
&
R_{v3}(z)
=
-\frac{1}{3}\frac{1}{(1-z)_+}+(1-z)\mylog{2-2z}+\frac{7}{6}z+\frac{1}{2}
\,,
\end{split}\\
%=========R v4==========
\begin{split}
&
R_{v4}(z)
=
\mylog{2-2z}-\frac{3}{4}z^2+\frac{3}{2}z-\frac{47}{60}
\,,
\end{split}\\
%=========R v5==========
\begin{split}
&
R_{v5}(z)
=
(1-z)\mylog{2-2z}-\frac{3}{4}z^2+\frac{109}{120}z+\frac{41}{120}
\,,
\end{split}\\
%=========R v6==========
\begin{split}
&
R_{v6}(z)
=
-\frac{5}{6}\frac{1}{(1-z)_+}+(1-z)\mylog{2-2z}+\frac{67}{60}z+\frac{1}{20}
\,,
\end{split}\\
%=========R v7==========
\begin{split}
&
R_{v7}(z)
=
\left(\frac{\text{ln}(2-2z)}{1-z}\right)_+ + \frac{1}{6}\frac{1}{(1-z)_+}
\,,
\end{split}\\
%=========R v8==========
\begin{split}
&
R_{v8}(z)
=
\left(\frac{\text{ln}(2-2z)}{1-z}\right)_+ -\frac{1}{30}\frac{1}{(1-z)_+} -\frac{3}{4}(1-z)
\,,
\end{split}\\
%========= c =============
\begin{split}
c
=
N_c^2(3+4\logtwo)+1,
\end{split}\\
%========= c_v =============
\begin{split}
c_{1}
=
-4\,N_c^2\big[(\logtwo)^2 +\logtwo -1\big]-4,
\end{split}
%====== End ===========
\end{align}
%

%%%%%%%%%%%%%%%%%%%%%%%%%%%%%%%%%%
%                                                                                                       %
%                       NLO results for a                                                     %
%                                                                                                      %
%%%%%%%%%%%%%%%%%%%%%%%%%%%%%%%%%%

\subsection{$P$-wave NLO results with an initial axial-vector $\cc$-state}

Here, we list results of NLO short-distance contributions to the FFs from an axial-vector pQCD $\cc$-state.

\begin{align}
%==========A1  to  1P1[1]===========
\begin{split}\label{eq:a1to1p11NLO}
&\hat{d}^{\text{(1)}}_{\aone\to \CSaPa}
=-\frac{1}{12}C_F\delta(1-z)
\Big\{
\frac{\DeltaPPZero}{4}(\LogUV-4)
\\
&\hspace{2cm}
+{\tilde{V}'}_{va}(\zeta_1,\zeta_2)(\LogUV-\frac{2}{3})
+{V_{a1}'}(\zeta_1,\zeta_2)
\Big\}
,
\end{split}\\
%==========A1  to  3P1[1]===========
\begin{split}\label{eq:a1to3p11NLO}
&\hat{d}^{\text{(1)}}_{\aone\to \CScPa}
=\frac{1}{6 }C_F\delta(1-z)
\Big\{
\frac{3}{4}\,\DeltaZero(\LogUV+\frac{2}{3})
\\
&\hspace{2cm}
+{\tilde{V}}_{va}(\zeta_1,\zeta_2)(\LogUV+\frac{4}{3})
+{V_{a2}}(\zeta_1,\zeta_2)
\Big\}
,
\end{split}\\
%==========A1  to  1P1[8]===========
\begin{split}\label{eq:a1to1p18NLO}
&\hat{d}^{\text{(1)}}_{\aone\to \COaPa}
=\frac{z}{12 }\frac{C_F}{(N_c^2-1)}
\Big\{
2\,\DeltaZero\,\delta(1-z)(-\LogIR+2\logtwo+\frac{1}{3})
\\
&\hspace{2cm}
+\big[\frac{\DPlusPPOne}{(1-z)} +\DPlusPOne +\DPlusOne(1-z)\big] (\frac{1}{2}\LogUV-\frac{1}{3})
\\
&\hspace{2cm}
-\frac{\DPlusPPOne}{(1-z)}  R_{a1}(z)- \frac{\DPlusPOne}{(1-z)} R_{a2}(z)-\DPlusOne \,R_{a3}(z)
\Big\}
,
\end{split}\\
%==========A1  to  3P0[8]===========
\begin{split}\label{eq:a1to3p08NLO}
&\hat{d}^{\text{(1)}}_{\aone\to \COcPz}
=\frac{1}{24 }\frac{C_F}{(N_c^2-1)}z(1-z)\big[\DMinusPPOne +\DMinusPOne +\DMinusOne\big]
(\LogUV-2\mylog{2-2z}-3)
,
\end{split}\\
%==========A1  to  3P1[8]===========
\begin{split}\label{eq:a1to3p18NLO}
&\hat{d}^{\text{(1)}}_{\aone\to \COcPa}
=\frac{z}{6}\frac{C_F}{(N_c^2-1)}\,
\Big\{
\frac{\DMinusOne}{(1-z)} (\frac{1}{2}\LogUV+\frac{2}{3})
\\
&\hspace{2cm}
-\DMinusPPOne R_{a4}(z)-\DMinusPOne R_{a5}(z)
-\frac{\DMinusOne}{(1-z)} R_{a6}(z)
\Big\}
,
\end{split}\\
%==========A1  to  3P2[8]===========
\begin{split}\label{eq:a1to3p28NLO}
&\hat{d}^{\text{(1)}}_{\aone\to \COcPb}
=\frac{1}{30}\frac{C_F}{(N_c^2-1)}z\,(1-z)
\Big\{
\big[\DMinusPPOne +\DMinusPOne +\DMinusOne\big] (\frac{1}{2}\LogUV-\frac{8}{15})
\\
&\hspace{2cm}
-
\DMinusPPOne \big[\mylog{2-2z}+\frac{13}{60}\big]
-\DMinusPOne \big[\mylog{2-2z}+\frac{161}{120}\big]
\\
&\hspace{2cm}
-\DMinusOne \big[\mylog{2-2z}+\frac{13}{60}\big]
\Big\}
,
\end{split}\\
%==========A8  to  1P1[8]===========
\begin{split}\label{eq:a8to1p18NLO}
&
\hat{d}^{\text{(1)}}_{\aeight \to \COaPa}
=
\frac{z}{12}
\frac{C_F}{(N_c^2-1)^2}
\Big\{
\delta(1-z)
\Big[
{(N_c^2-4)}\DeltaZero\Big(-\LogIR+2\logtwo+\frac{1}{3}\Big)
\\
&\hspace{2cm}
+\frac{1}{4}\DeltaPPZero\,\Big(c \times \LogUV+c_{1}\Big)
+\tilde{V}'_{va}(\zeta_1,\zeta_2)\big(\LogUV-\frac{2}{3}\,\big)
+V'_{a1}(\zeta_1,\zeta_2)
\Big]
\\
&\hspace{2cm}
+\Big[
 \frac{\DMinusPPEight}{(1-z)_+}+{\DMinusPEight}+{\DMinusEight} (1-z)
\Big]\big(\, \frac{1}{2}\LogUV-\frac{1}{3}\,\big)
\\
&\hspace{2cm}
-
\DMinusPPEight\,R_{a7}(z)-\frac{\DMinusPEight}{(1-z)}R_{a2}(z)-\DMinusEight R_{a3}(z)
\Big\}
,
\end{split}\\
%==========A8  to  3P0[8]===========
\begin{split}\label{eq:a8to3p08NLO}
&\hat{d}^{\text{(1)}}_{\aeight\to \COcPz}
=\frac{1}{24 }\frac{C_F}{(N_c^2-1)^2}z(1-z)\big[\DPlusPPEight+\DPlusPEight+\DPlusEight\big]
\\
&\hspace{2cm}
\times
(\LogUV-2\mylog{2-2z}-3)
,
\end{split}\\
%==========A8  to  3P1[8]===========
\begin{split}\label{eq:a8to3p18NLO}
&
\hat{d}^{\text{(1)}}_{\aeight \to \COcPa}
=
-\frac{z}{6}
\frac{C_F}{(N_c^2-1)^2}
\Big\{
\delta(1-z)
\Big[\,
\frac{3}{4}\DeltaZero \Big(\tilde{c} \times \LogUV+c_{2}\Big)
\\
&\hspace{2cm}
+\tilde{V}_{va}(\zeta_1,\zeta_2)\big( \LogUV+\frac{4}{3}\,\big)
+\,V_{a2}(\zeta_1,\zeta_2)
\Big]
-
\frac{\DPlusEight}{(1-z)_+}\big(\, \frac{1}{2}\LogUV+\frac{2}{3}\,\big)
\\
&\hspace{2cm}
+
\DPlusPPEight\,R_{a4}(z)+\DPlusPEight R_{a5}(z)+\DPlusEight R_{a8}(z)
\Big\}
,
\end{split}\\
%==========A8  to  3P2[8]===========
\begin{split}\label{eq:a8to3p28NLO}
&\hat{d}^{\text{(1)}}_{\aeight\to \COcPb}
=\frac{1}{30  }\frac{C_F}{(N_c^2-1)^2}z\,(1-z)
\Big\{
\big[\DPlusPPEight+\DPlusPEight +\DPlusEight\big] (\frac{1}{2}\LogUV-\frac{8}{15})
\\
&\hspace{2cm}
-
\DPlusPPEight \big[\mylog{2-2z}+\frac{13}{60}\big]
-\DPlusPEight \big[\mylog{2-2z}+\frac{161}{120}\big]
\\
&\hspace{2cm}
-\DPlusEight \big[\mylog{2-2z}+\frac{13}{60}\big]
\Big\}
,
\end{split}\\
%==========s8  to  s[1]===========
\begin{split}\label{eq:s8tos1NLO}
\hat{d}^{\text{(1)}}_{a^{[8]} \to \state{{2S+1}}{L}{J}{1}}
=\hat{d}^{\text{(1)}}_{a^{[1]} \to \state{{2S+1}}{L}{J}{8}},
\end{split}
%========The End==============
\end{align}
where, again, the dependence on $z$, $\zeta_1$, $\zeta_2$ and $\mu_F$ in the last equation is suppressed.
$\tilde{V}$, $V$, $R$ and $c$ above are defined as
%
%======== Aux functions =============
\begin{align}
%=========\V_{va}==========
\begin{split}
{\tilde{V}}_{va}(\zeta_1,\zeta_2)
=
\delta(\zeta_2)\Big\{
\Plusa-\frac{1}{2}\Plusz{\zeta_1+1}
\Big\}
+(\zeta_1 \leftrightarrow \zeta_2)
,
\end{split}\\
%=========V'_{a1}==========
\begin{split}
&{V'_{a1}}(\zeta_1,\zeta_2)
=
\delta'(\zeta_2)
\Big\{
\Minusc
-\MinusLogb+\frac{5}{3}\Minusb-3\Minusa
\\
&\hspace{2cm}
+\frac{1}{2}\Minusz{(\zeta_1+1)\mylog{\zeta_1^2}}
+\frac{7}{6}\Minusz{\zeta_1-\frac{5}{7}}
\Big\}
+(\zeta_1 \leftrightarrow \zeta_2)
,
\end{split}\\
%=========V_{a2}==========
\begin{split}
&{V_{a2}}(\zeta_1,\zeta_2)
=
\delta(\zeta_2)
\Big\{
\frac{1}{2}\Plusb
-\PlusLoga-\frac{7}{3}\Plusa+\frac{1}{2}\Plusz{(\zeta_1+1)\text{ln}(\zeta_1^2)}
\\
&\hspace{2cm}
+\frac{1}{6}\Plusz{\zeta_1+10}
\Big\}
+(\zeta_1 \leftrightarrow \zeta_2)
,
\end{split}\\
%=========R a1==========
\begin{split}
&
R_{a1}(z)
=
\mylog{2-2z}+\frac{1}{6}
\,,
\end{split}\\
%=========R a2==========
\begin{split}
&
R_{a2}(z)
=
(1-z)\mylog{2-2z}-\frac{1}{6}z+\frac{2}{3}
\,,
\end{split}\\
%=========R a3==========
\begin{split}
&
R_{a3}(z)
=
-\frac{1}{(1-z)_+}+(1-z)\mylog{2-2z}+\frac{5}{6}z+\frac{7}{6}
\,,
\end{split}\\
%=========R a4==========
\begin{split}
&
R_{a4}(z)
=
-\frac{1}{4}\,(1-z)
\,,
\end{split}\\
%=========R a5==========
\begin{split}
&
R_{a5}(z)
=
-\frac{3}{8}\,(1-z)
\,,
\end{split}\\
%=========R a6==========
\begin{split}
&
R_{a6}(z)
=
\mylog{2-2z}+\frac{5}{4}z^2-3z+\frac{35}{12}
\,,
\end{split}\\
%=========R a7==========
\begin{split}
&
R_{a7}(z)
=
\left(\frac{\text{ln}(2-2z)}{1-z}\right)_+ + \frac{1}{6}\frac{1}{(1-z)_+}
\,,
\end{split}\\
%=========R a8==========
\begin{split}
&
R_{a8}(z)
=
\left(\frac{\text{ln}(2-2z)}{1-z}\right)_+ + \frac{7}{6}\frac{1}{(1-z)_+} -\frac{5}{4}z+\frac{7}{4}
\,,
\end{split}\\
%========= \tilde{c} =============
\begin{split}
\tilde{c}
=
1 - N_c^2(1+\frac{4}{3}\logtwo),
\end{split}\\
%========= c_2 ==========
\begin{split}
&
c_{2}
=
\frac{4}{3}N_c^2\big[(\logtwo)^2 +\logtwo -1\big]+\frac{2}{3},
\end{split}
%====== End ===========
\end{align}
%

%%%%%%%%%%%%%%%%%%%%%%%%%%%%%%%%%%
%                                                                                                       %
%                       NLO results for t                                                     %
%                                                                                                      %
%%%%%%%%%%%%%%%%%%%%%%%%%%%%%%%%%%

\subsection{P-wave NLO results with an initial tensor $\cc$-state}

Here, we list results of NLO short-distance contributions to the FFs for a tensor pQCD $\cc$-state to a non-relativistic $P$-wave $\cc$-state.

\begin{align}
%==========T1  to  1P1[1]===========
\begin{split}\label{eq:t1to1p11NLO}
&\hat{d}^{\text{(1)}}_{\tone\to \CSaPa}
=\frac{1}{12 }C_F\delta(1-z)
\Big\{
\frac{3}{4}\DeltaZero(\LogUV+\frac{2}{3})
\\
&\hspace{2cm}
+{\tilde{V}}_{t}(\zeta_1,\zeta_2)(\LogUV+\frac{4}{3})
+{V_{t1}}(\zeta_1,\zeta_2)
\Big\}
,
\end{split}\\
%==========T1  to  3P1[1]===========
\begin{split}\label{eq:t1to3p11NLO}
&\hat{d}^{\text{(1)}}_{\tone\to \CScPa}
=-\frac{1}{24 }C_F\delta(1-z)
\Big\{
\frac{\DeltaPPZero}{4}(\LogUV-4)
\\
&\hspace{2cm}
+{\tilde{V}'}_{t}(\zeta_1,\zeta_2)(\LogUV-\frac{5}{3})
+{V_{t2}'}(\zeta_1,\zeta_2)
\Big\}
,
\end{split}\\
%==========T1  to  3P2[1]===========
\begin{split}\label{eq:t1to3p21NLO}
&\hat{d}^{\text{(1)}}_{\tone\to \CScPb}
=-\frac{1}{40  }C_F\delta(1-z)
\Big\{
\frac{\DeltaPPZero}{4}(\LogUV-4)
\\
&\hspace{2cm}
+{\tilde{V}'}_{t}(\zeta_1,\zeta_2)(\LogUV-\frac{7}{5})
+{V_{t3}'}(\zeta_1,\zeta_2)
\Big\}
,
\end{split}\\
%==========T1  to  1P1[8]===========
\begin{split}\label{eq:t1to1p18NLO}
&\hat{d}^{\text{(1)}}_{\tone\to \COaPa}
=\frac{1}{12 }\frac{C_F}{(N_c^2-1)}\frac{z}{(1-z)}\DMinusOne
\Big\{
(z^2-2z+2) (\frac{1}{2}\LogUV+\frac{2}{3})
-R_{t1}(z)
\Big\}
,
\end{split}\\
%==========T1  to  3P0[8]===========
\begin{split}\label{eq:t1to3p08NLO}
&\hat{d}^{\text{(1)}}_{\tone\to \COcPz}
=\frac{z}{24}\frac{C_F}{(N_c^2-1)}
\Big\{
\frac{4}{3}\DeltaZero\,\delta(1-z)(-\LogIR+2\logtwo+\frac{1}{2})
\\
&\hspace{2cm}
+\DPlusPPOne(1-z) +\frac{1}{2} \DPlusPOne (4-3z)+\frac{2}{3}\DPlusOne \big(\frac{1}{(1-z)_+}-4z+5\big)
\Big\}
,
\end{split}\\
%==========T1  to  3P1[8]===========
\begin{split}\label{eq:t1to3p18NLO}
&\hat{d}^{\text{(1)}}_{\tone\to \COcPa}
=\frac{z}{24}\frac{C_F}{(N_c^2-1)}
\Big\{
\frac{4}{3}\DeltaZero\,\delta(1-z)(-\LogIR+2\logtwo+\frac{1}{4})
\\
&\hspace{1.6cm}
+\big[\frac{\DPlusPPOne}{2(1-z)}(z^2-2z+2)+\frac{\DPlusPOne}{2} (2-z)+\DPlusOne(1-z)\big] (\LogUV-\frac{5}{3})
\\
&\hspace{1.6cm}
-\frac{\DPlusPPOne}{(1-z)}  R_{t2}(z)- \frac{\DPlusPOne}{(1-z)} R_{t3}(z)-\DPlusOne \,R_{t4}(z)
\Big]
\Big\}
,
\end{split}\\
%==========t1  to  3P2[8]===========
\begin{split}\label{eq:v1to3p28NLO}
&\hat{d}^{\text{(1)}}_{\tone\to \COcPb}
=\frac{z}{40 }\frac{C_F}{(N_c^2-1)}
\Big\{
\frac{20}{9}\DeltaZero\,\delta(1-z)(-\LogIR+2\logtwo+\frac{7}{20})
\\
&\hspace{1.6cm}
+\big[\frac{\DPlusPPOne}{2(1-z)}(z^2-2z+2)+\frac{\DPlusPOne}{2} (2-z)+\DPlusOne(1-z)\big] (\LogUV-\frac{7}{5})
\\
&\hspace{1.6cm}
-\frac{\DPlusPPOne}{(1-z)}  R_{t5}(z)- \frac{\DPlusPOne}{(1-z)} R_{t6}(z)-\DPlusOne \,R_{t7}(z)
\Big]
\Big\}
,
\end{split}\\
%==========T8  to  1P1[8]===========
\begin{split}\label{eq:t8to1p18NLO}
&\hat{d}^{\text{(1)}}_{\teight\to \COaPa}
=
-\frac{z}{12}
\frac{C_F}{(N_c^2-1)^2}
\Big\{
\delta(1-z)
\Big[\,
\frac{3}{4}\DeltaZero \Big(\tilde{c} \times \LogUV+c_{2}\Big)
\\
&\hspace{2cm}
+\tilde{V}_{t}(\zeta_1,\zeta_2)\big(\LogUV+\frac{4}{3}\big)
+V_{t1}(\zeta_1,\zeta_2)
\Big]
\\
&\hspace{2cm}
-\DPlusEight\frac{(z^2-2z+2)}{(1-z)_+}\big(\,\frac{1}{2}\LogUV+\frac{2}{3}\big)
+\DPlusEight\,R_{t8}(z)
\Big\}
,
\end{split}\\
%==========T8  to  3P0[8]===========
\begin{split}\label{eq:t8to3p08NLO}
&
\hat{d}^{\text{(1)}}_{\teight\to \COcPz}
=\frac{z}{24  }\frac{C_F}{(N_c^2-1)^2}
\Big\{
\frac{2(N_c^2-4)}{3}\DeltaZero\,\delta(1-z)(-\LogIR+2\logtwo+\frac{1}{2})
\\
&\hspace{2cm}
+\DMinusPPEight (1-z) +\frac{1}{2}\DMinusPEight (4-3z)+\frac{2}{3}\DMinusEight \big(\frac{1}{(1-z)_+}-4z+5\big)
\Big\}
,
\end{split}\\
%==========T8  to  3P1[8]===========
\begin{split}\label{eq:t8to3p18NLO}
&\hat{d}^{\text{(1)}}_{\teight\to \COcPa}
=
\frac{z}{24}
\frac{C_F}{(N_c^2-1)^2}
\Big\{
\delta(1-z)
\Big[\,
 \frac{2}{3}(N_c^2-4)\DeltaZero\Big(-\LogIR+2\logtwo+\frac{1}{4}\Big)
\\
&\hspace{1.6cm}
+\frac{1}{4}\DeltaPPZero \Big(c \times \LogUV+c_{1}\Big)
+\tilde{V}_{t}'(\zeta_1,\zeta_2)\big( \LogUV-\frac{5}{3}\,\big)
+V'_{t2}(\zeta_1,\zeta_2)
\Big]
\\
&\hspace{1.6cm}
+
\Big[
\DMinusPPEight \frac{ (z^2-2z+2)}{2(1-z)_+}+\frac{\DMinusPEight}{2}(2-z)+\DMinusEight (1-z)
\Big]\big( \LogUV-\frac{5}{3}\,\big)
\\
&\hspace{1.6cm}
-
\DMinusPPEight\,R_{t9}(z)-\frac{\DMinusPEight}{(1-z)}R_{t3}(z)-\DMinusEight R_{t4}(z)
\Big\}
,
\end{split}\\
%==========T8  to  3P2[8]===========
\begin{split}\label{eq:t8to3p28NLO}
&
\hat{d}^{\text{(1)}}_{\teight \to \COcPb}
=
\frac{z}{40}
\frac{C_F}{(N_c^2-1)^2}
\Big\{
\delta(1-z)
\Big[\,
\frac{10}{9}(N_c^2-4)\DeltaZero\Big(-\LogIR+2\logtwo+\frac{7}{20}\Big)
\\
&\hspace{1.6cm}
+\frac{1}{4}\DeltaPPZero \Big(c \times \LogUV+c_{1}\Big)
+\tilde{V}_{t}'(\zeta_1,\zeta_2)\big( \LogUV-\frac{7}{5}\,\big)
+V'_{t3}(\zeta_1,\zeta_2)
\Big]
\\
&\hspace{1.6cm}
+\Big[
\DMinusPPEight \frac{ (z^2-2z+2)}{2(1-z)_+}+\frac{\DMinusPEight}{2}(2-z)+\DMinusEight (1-z)
\Big]\big( \LogUV-\frac{7}{5}\,\big)
\\
&\hspace{1.6cm}
-
\DMinusPPEight\,R_{t10}(z)-\frac{\DMinusPEight}{(1-z)}R_{t6}(z)-\DMinusEight R_{t7}(z)
\Big\}
,
\end{split}\\
%==========s8  to  s[1]===========
\begin{split}\label{eq:s8tos1NLO}
\hat{d}^{\text{(1)}}_{t^{[8]} \to \state{{2S+1}}{L}{J}{1}}
=\hat{d}^{\text{(1)}}_{t^{[1]} \to \state{{2S+1}}{L}{J}{8}},
\end{split}
%========The End==============
\end{align}
where the dependence on $z$, $\zeta_1$, $\zeta_2$, and $\mu_F$ in the last equation is suppressed.
$\tilde{V}$, $V$, $R$ and $c$ above are defined as
%
%======== Aux functions =============
\begin{align}
%=========\tilde V_{t}==========
\begin{split}
{\tilde{V}}_{t}(\zeta_1,\zeta_2)
=
\delta(\zeta_2)\Big\{
\Plusa-\Plusz{1}
\Big\}
+(\zeta_1 \leftrightarrow \zeta_2)
,
\end{split}\\
%=========\tilde V'_{t}==========
\begin{split}
{\tilde{V}'}_{t}(\zeta_1,\zeta_2)
=
\delta'(\zeta_2)\Big\{
\Minusb-\Minusz{1}
\Big\}
+(\zeta_1 \leftrightarrow \zeta_2)
,
\end{split}\\
%=========V_{t1}==========
\begin{split}
&{V_{t1}}(\zeta_1,\zeta_2)
=
\delta(\zeta_2)
\Big\{
\frac{1}{2}\Plusb-\PlusLoga-\frac{7}{3}\Plusa+\Plusz{\mylog{\zeta_1^2}}+\frac{11}{6}\Plusz{1}
\Big\}
\\
&\hspace{2cm}
+(\zeta_1 \leftrightarrow \zeta_2)
,
\end{split}\\
%=========V'_{t2}==========
\begin{split}
&{V'_{t2}}(\zeta_1,\zeta_2)
=
\delta'(\zeta_2)
\Big\{
\Minusc
-\MinusLogb+\frac{8}{3}\Minusb-2\Minusa
\\
&\hspace{2cm}
+\Minusz{\mylog{\zeta_1^2}}
-{\frac{5}{3}}\Minusz{1}
\Big\}
+(\zeta_1 \leftrightarrow \zeta_2)
,
\end{split}\\
%=========V'_{t3}==========
\begin{split}
&{V'_{t3}}(\zeta_1,\zeta_2)
=
\delta'(\zeta_2)
\Big\{
\Minusc
-\MinusLogb+\frac{12}{5}\Minusb-4\Minusa
\\
&\hspace{2cm}
+\Minusz{\mylog{\zeta_1^2}}
+{\frac{3}{5}}\Minusz{1}
\Big\}
+(\zeta_1 \leftrightarrow \zeta_2)
,
\end{split}\\
%=========R t1==========
\begin{split}
&
R_{t1}(z)
=
(z^2-2z+2)\mylog{2-2z}+\frac{5}{3}z^2-\frac{23}{6}z+\frac{10}{3}
,
\end{split}\\
%=========R t2==========
\begin{split}
&
R_{t2}(z)
=
(z^2-2z+2)\mylog{2-2z}+\frac{1}{6}z^2-\frac{1}{3}z-\frac{1}{6}
\,,
\end{split}\\
%=========R t3==========
\begin{split}
&
R_{t3}(z)
=
(z^2-3z+2)\mylog{2-2z}+\frac{5}{12}z^2-\frac{1}{4}z+\frac{1}{3}
\,,
\end{split}\\
%=========R t4==========
\begin{split}
&
R_{t4}(z)
=
-\frac{2}{3}\frac{1}{(1-z)_+}z^2+\frac{7}{6}z+2(1-z)\mylog{2-2z}-\frac{1}{6}
\,,
\end{split}\\
%=========R t5==========
\begin{split}
&
R_{t5}(z)
=
(z^2-2z+2)\mylog{2-2z}-\frac{11}{30}z^2+\frac{22}{30}z-\frac{17}{30}
\,,
\end{split}\\
%=========R t6==========
\begin{split}
&
R_{t6}(z)
=
(z^2-3z+2)\mylog{2-2z}+\frac{11}{20}z^2-\frac{79}{60}z+\frac{76}{60}
\,\end{split}\\
%=========R t7==========
\begin{split}
&
R_{t7}(z)
=
-\frac{10}{9}\frac{1}{(1-z)_+}+\frac{31}{90}z+2(1-z)\mylog{2-2z}+\frac{49}{90}
\,,
\end{split}\\
%=========R t8==========
\begin{split}
&
R_{t8}(z)
=
\left(\frac{\text{ln}(2-2z)}{1-z}\right)_+ +\frac{7}{6}\frac{1}{(1-z)_+} +(1-z)\mylog{2-2z}
-\frac{5}{3}z+\frac{13}{6}
\,,
\end{split}\\
%=========R t9==========
\begin{split}
&
R_{t9}(z)
=
\left(\frac{\text{ln}(2-2z)}{1-z}\right)_+ -\frac{1}{3}\frac{1}{(1-z)_+} +(1-z)\mylog{2-2z}
+\frac{1}{6}(1-z)
\,,
\end{split}\\
%=========R t9==========
\begin{split}
&
R_{t10}(z)
=
\left(\frac{\text{ln}(2-2z)}{1-z}\right)_+ -\frac{1}{5}\frac{1}{(1-z)_+} +(1-z)\mylog{2-2z}
-\frac{11}{30}(1-z)
\,.
\end{split}
%====== End ===========
\end{align}

Note, finally, that all these short-distance contributions to the FFs, so as the FFs, are
invariant under the transformation ($\zeta_1 \to -\zeta_1$, $\zeta_2 \to -\zeta_2$) and
the exchange $\zeta_1 \leftrightarrow \zeta_2$, including the crossing exchange
($\zeta_1 \to -\zeta_2$, $\zeta_2 \to -\zeta_1$), which are the features derived from
the general symmetries of QCD.

%%%%%%%%%%%%%%%%%%%%%%%%%%%%%%%%%%
%                                                                                                       %
%                      compare with others                                                 %
%                                                                                                      %
%%%%%%%%%%%%%%%%%%%%%%%%%%%%%%%%%%

\subsection{Comparison with Other Calculations}

Almost at the same time, color singlet to color singlet processes, means Eqs. (\ref{eq:v1to3p01NLO}, \ref{eq:v1to3p21NLO}, \ref{eq:a1to1p11NLO}, \ref{eq:a1to3p11NLO}, \ref{eq:t1to1p11NLO}, \ref{eq:t1to3p11NLO}, \ref{eq:t1to3p21NLO}), were also calculated independently in Ref.~\cite{Wang:2013ywc} in the terminology of distribution amplitude. We find that Eq.~(\ref{eq:v1to3p01NLO}, \ref{eq:v1to3p21NLO}, \ref{eq:a1to1p11NLO}, \ref{eq:t1to3p21NLO}) are consistent with their results. Eq.~(\ref{eq:a1to3p11NLO}) cannot be compared with their result due to we use a different $\gamma_5$ scheme from theirs. By using their $\gamma_5$ scheme, we can indeed reproduce their results. The $\gamma_5$ scheme used in our calculation is discussed in Appendix B of the companion paper \cite{Swave}. Eq.~(\ref{eq:t1to1p11NLO}, \ref{eq:t1to3p11NLO}) cannot be compared directly with the corresponding results in Ref.~\cite{Wang:2013ywc} because the two calculations use different projection operators. By adopting the projection operators used in Ref.~\cite{Wang:2013ywc} for these two processes, we can reproduce their results.

%%%%%%%%%%%%%%%%%%%%%%%%%%%%%%%%%
%                                                   						  %
%                                   References						  %
%  											   	  %
%%%%%%%%%%%%%%%%%%%%%%%%%%%%%%%%%

\providecommand{\href}[2]{#2}\begingroup\raggedright\endgroup

%======================= The End =================================
\end{document}